# Approximate Bayesian Computation with Deep Learning and Conformal prediction

Meïli Baragatti[1], Céline Casenave[2], Bertrand Cloez[2], David Métivier[2], and Isabelle Sanchez[2]

[1]MISTEA, Université de Montpellier, INRAE, Institut Agro, Montpellier, France

[2]MISTEA, Université de Montpellier, INRAE, Montpellier, France

June 19, 2025

**Abstract**

Approximate Bayesian Computation (ABC) methods are commonly used to approximate posterior distributions in models with unknown or computationally intractable likelihoods. Classical ABC methods are based on nearest neighbor type algorithms and rely on the choice of so-called summary statistics, distances between datasets and a tolerance threshold. Recently, methods combining ABC with more complex machine learning algorithms have been proposed to mitigate the impact of these "user-choices". In this paper, we propose the first, to our knowledge, ABC method completely free of summary statistics, distance, and tolerance threshold. Moreover, in contrast with usual generalizations of the ABC method, it associates a confidence interval (having a proper frequentist marginal coverage) with the posterior mean estimation (or other moment-type estimates).

Our method, named ABCD-Conformal, uses a neural network with Monte Carlo Dropout to provide an estimation of the posterior mean (or other moment type functionals), and conformal theory to obtain associated confidence sets. Efficient for estimating multidimensional parameters and amortized, we test this new method on four different applications and compare it with other ABC methods in the literature. Supplementary materials are available online (appendix, source codes, notebooks, datasets).

# 1 Introduction

In many situations, the likelihood function plays a key role in both frequentist or Bayesian inference. However, it is not unusual that such likelihood is difficult to work with either because no closed form expression exists, e.g., the corresponding statistical model is too complex, or because its computational cost is prohibitive. For example, the first situation occurs with max-stable processes [de Haan, 1984] or Gibbs random fields [Besag, 1974; Grelaud et al., 2009], while the second case frequently occurs with mixture models [Simola et al., 2021].

To overcome the intractability of the likelihood function, several approaches have been explored, among which the use of composite likelihoods [Lindsay, 1988; Varin et al., 2011], that is linear combinations of smaller order log-likelihoods. The use of composite likelihoods in place of the (full) likelihood has been shown to be very efficient, but their application is typically limited to the frequentist framework as Bayes' theorem does not apply with composite likelihoods (although some work exists to obtain what we shall call "relevant composite posterior distributions" [Ribatet et al., 2012; Stoehr and Friel, 2015]). Other likelihood-free approaches, called Simulation Based Inference (SBI), use neural conditional density estimation to learn a parametric approximation of the likelihood [Papamakarios et al., 2019] or of the exact posterior [Papamakarios and Murray, 2016]. Both of these can also be approximated using nonparametric approaches; see Gloeckler et al. [2024] who propose a new amortized Bayesian inference method using neural networks.

In this paper, we will restrict our attention to parameter inference in the Bayesian framework and in particular to Approximate Bayesian Computation (ABC), which is one of the most popular likelihood-free parameter inference method. Indeed, ABC can be an efficient alternative to overcome the situation of intractable likelihood functions [Marin et al., 2012]. Within the ABC framework, the need for computing the likelihood function is substituted by the need to be able to simulate from the model under consideration. The first proposed ABC methods enabled to obtain an approximation of the posterior distribution of interest, from simulations [Pritchard et al., 1999]. The simplest sampler is an accept-reject algorithm, which will be referred to as "standard ABC" in the following. This vanilla sampler can be quite inefficient, as numerous draws may be rejected before one is accepted. Alternative samplers have therefore been proposed in a search for computational efficiency, using MCMC or filtering algorithms (see Marjoram et al. [2003], Sisson et al. [2007], Baragatti and Pudlo [2014] and references therein). In particular, the Sequential Monte Carlo ABC (ABC-SMC, Sisson et al. [2007]; Beaumont et al. [2009]; Del Moral et al. [2012]) has become the gold standard among ABC methods. Even though more efficient than the standard ABC, these early ABC samplers are generally difficult to use when simulating complex systems or inferring high-dimensional parameters. This is why, with the increasing complexity of the data and models used nowadays, ABC samplers remain an active field of research. For instance, in order to improve the ABC-SMC sampler, guided approaches have been proposed with the goal of rapidly targeting the region of the parameter space that is compatible with the observed data [Picchini and Tamborrino, 2024]. However, for most of these ABC methods, the degree of approximation of the posterior distribution depends strongly on the choices of the tolerance threshold, the distance and of the so-called summary statistics used during the ABC estimation. Unfortunately, identifying the most relevant summary statistics is problem dependent and in many situations the statistics are chosen by the practitioner, even if several studies provide guidance in this choice (see for instance Joyce and Marjoram [2008], McKinley et al. [2009], Blum et al. [2012], Fearnhead and Prangle [2012] among others).

Recently, several approaches managed to avoid using the tolerance level or to mitigate the need of identifying relevant summary statistics. Most of these approaches combine the ABC framework with machine learning methods, such as random forests or neural networks. For instance, Raynal et al. [2018] propose an automatic selection of relevant summary statistics. They propose to consider a large set of potentially relevant summary statistics and use the random forest framework as a black box to get some point estimates of the posterior distribution given these summary statistics. They do not need any tolerance level, like the previously cited Simulation Based Inference (SBI) methods [Papamakarios and Murray, 2016; Papamakarios et al., 2019], which are called "$\epsilon$-free" by their authors and use neural networks. Another approach of interest is the one of Akesson et al. [2022], which uses the estimated posterior mean as a summary statistic, this posterior mean being estimated through a Convolutional Neural Network (CNN). To summarise, since the emergence of ABC methods, a wide variety of methods have been proposed, from the standard "basic" ABC to methods using random forests or neural networks.

To quantify the uncertainties, Bayesian methods provide posterior credible intervals from the posterior distribution, through posterior quantiles or highest posterior density regions. This can be justified because, in many simple parametric models, the Bernstein-von Mises theorem ensures that posterior credible intervals are asymptotically as well, reconciling the frequentist and Bayesian approaches. However, these credible intervals are not necessarily relevant in practice. Indeed, note for instance that in nonparametric and high-dimensional models, research is still ongoing to understand better the link between the coverage of credible intervals and the coverage of frequentist confidence intervals, see Rousseau and Szabo [2016]; Datta et al. [2000]; Hoff [2023] among others. Indeed, the posterior coverage probability of a Bayesian credible interval does not always match with the corresponding frequentist coverage probability. Quite often, they can even diverge in nontrivial ways, see for instance Wasserman [2011]. In this case, we say that the *probability matching criterion* is not verified. This criterion can be hard to check when the exact posterior distribution is known, and even more so when the posterior is approximated [Frazier et al., 2018], which is the case of most applications requiring an ABC method. Another drawback of using posterior credible intervals in ABC methods is that they only take into account the inherent randomness of the model when estimating the prediction uncertainty. Nevertheless, by definition of the method, the a posteriori distribution is approximated, so there is an error due to the method to take into account.

**Our contribution.** In this paper, we propose a new ABC implementation that does not require summary statistics, tolerance threshold and distance. Our method produces point estimates of the posterior distribution e.g., mean, variance, quantiles, without aiming to approximate the whole posterior distribution. It also provides associated confidence sets of the estimates. The estimates are obtained using neural networks with Monte Carlo Dropout, while the confidence sets are obtained through conformal prediction. Our ABC estimation method only requires 1) a sampling function, 2) a neural network (NN) whose architecture is tailored for the problem of interest e.g., CNN for images, GNN for data with a graph structure, RNN for sequential data. Furthermore, to obtain confidence intervals we need 3) a confidence level $\delta$. While 1) is common to all ABC methods, requirement 2) is more original, but still easy to implement because well-known architectures can be used with minimal changes, i.e., the inclusion of Dropout (or concrete Dropout) layers. Note that the choice of the architecture in 2) is not directly comparable to the choice of a summary statistics because the same architecture (with only minor changes) can be reused for very different applications. Another originality of this work is that it directly produces confidence sets with proper marginal coverage, whereas usual ABC methods enable one to obtain credible sets using $\delta$-quantiles of the estimated posterior distribution.

We test the proposed ABCD-Conformal algorithm in four different problems. We compare the results against the standard ABC, the ABC-SMC, the ABC Random Forest (ABC-RF) of Raynal et al. [2018] and the ABC-CNN of Akesson et al. [2022]. We measure both the accuracy of the estimation and the frequentist coverage of confidence sets. We provide a link to the Git repository `https://forge.inrae.fr/mistea/codes_articles/abcdconformal` where are stored all the codes, instructions and notebooks to reproduce our results in Julia and R. The Julia package `ABCMethods.jl` [Métivier, 2025] was developed for this paper.

The paper is organized as follows. In section 2 we briefly present some existing ABC methods: standard ABC, ABC-SMC, ABC-CNN and ABC-RF. In section 3, we propose the ABCD-Conformal methodology, combining neural networks and conformal prediction in a likelihood-free framework. Then, we illustrate and compare all these methods on three examples in section 4, before a discussion in section 5.

# 2 A few approaches of Approximate Bayesian Computation

## 2.1 Standard ABC, vanilla sampler

Consider a parametric statistical model $\{f(x \mid \theta)\colon x \in \mathbb{R}^d, \theta \in \Theta\}$ on which we assume a prior distribution $\pi$ on the parameter $\theta$ leading to the Bayesian parametric model $\{f(\cdot \mid \theta), \pi\}$. Based on the sample $\mathbf{x} = (x_1, \ldots, x_n)$, the cornerstone of any Bayesian inference is the posterior distribution,

$$\pi(\theta \mid \mathbf{x}) = \frac{f(\mathbf{x} \mid \theta)\pi(\theta)}{\int_\Theta f(\mathbf{x} \mid \theta)\pi(\theta)\mathrm{d}\theta} \propto f(\mathbf{x} \mid \pi)\pi(\theta), \tag{1}$$

where $f(\mathbf{x} \mid \theta)$ denotes the likelihood function for the sample $\mathbf{x}$ and parameter $\theta$. In a setting where we assume that the likelihood is not tractable, the simple, standard ABC methodology was the first powerful algorithm to bypass this hurdle (see Pritchard et al. [1999] or Marin et al. [2012]) and, in one of its simplest form, is described in Algorithm 1. In Algorithm 1,

---

**Algorithm 1:** Pseudocode for the standard ABC sampler

---

**Input** : A Bayesian parametric model $\{f(\cdot \mid \theta), \pi\}$, a data sample $\mathbf{x}$, the size of the training set $N_{\text{train}}$, summary statistics $\eta : \mathbb{R}^d \to \mathbb{R}^l$, a (pseudo)-distance $d$ on $\mathbb{R}^l$, and a tolerance threshold $\alpha \in (0, 1]$.

**Output:** A set $\theta_1, \ldots, \theta_{[N_{\text{train}}\alpha]}$ whose distribution is approximately $\pi(\theta \mid \mathbf{x})$

```
1  Generation of a reference table :
2  for j ← 1 to N_train do
3  |  Draw θ_j ~ π;
4  |  Draw synthetic sample x_j = (x_{1,j}, ..., x_{d,j})^T from the model f(· | θ_j)
5  end
6  Summary statistics and distances :
7  for j ← 1 to N_train do
8  |  Compute the vector of summary statistics η(x_j);
9  |  Compute the (pseudo)-distance d_j = d(η(x), η(x_j))
10 end
11 Order these distances, i.e., d_(1) < d_(2) < ··· < d_(N_train);
12 Keep the θ_j corresponding to the ⌊N_train α⌋ smallest distances.
```

---

$d$ is typically a pseudo distance of the form:

$$
\begin{aligned}
d \colon \mathbb{R}^l \times \mathbb{R}^l &\longrightarrow \mathbb{R}_+ \\
(\eta(\mathbf{x}), \eta(\mathbf{x})) &\longmapsto \|\eta(\mathbf{x}) - \eta(\mathbf{x})\|,
\end{aligned}
$$

where $\eta$ summarizes the data and is called the vector of summary statistics. $\|\cdot\|$ is typically the usual Euclidean norm. This algorithm samples $\theta$ and $\mathbf{x}$ from the joint posterior density

$$
\pi_{\alpha,d}(\theta, \mathbf{x} \mid \eta(\mathbf{x})) = \frac{\pi(\theta) f(\mathbf{x} \mid \theta) \mathbf{1}_{\alpha,d}(\mathbf{x})}{\int \pi(\theta) f(\mathbf{x} \mid \theta) \mathbf{1}_{\alpha,d}(\mathbf{x}) d\mathbf{x} d\theta},
$$

where $\mathbf{1}_{\alpha,d}(\mathbf{x}) = \mathbf{1}\big[d\big(\eta(\mathbf{x}), \eta(\mathbf{x})\big) < d_{([\alpha N]+1)}\big]$. The approximate Bayesian computation posterior density is defined as:

$$
\pi_{\alpha,d}(\theta \mid \eta(\mathbf{x})) = \int \pi_{\alpha,d}(\theta, \mathbf{x} \mid \eta(\mathbf{x})) d\mathbf{x}
$$

The basic idea behind ABC is that using representative (enough) summary statistics $\eta$ coupled with a small (enough) tolerance level $\alpha$ should produce a good (enough) approximation to the posterior distribution; that is $\pi_{\alpha,d}(\theta \mid \eta(\mathbf{x})) \approx \pi(\theta \mid \mathbf{x})$. Different theoretical results have validated this approximation. Frazier et al. [2018] prove that the posterior distribution concentrates on sets containing the true parameter under general assumptions and with some rates of convergence. Fearnhead and Prangle [2012] obtain the best theoretical summary statistics to perform this approximation, under the hypothesis of a quadratic loss.

Nevertheless, for the algorithm to give a valid approximation of the true posterior distribution, not only $\pi_{\alpha,d}$ has to be close to $\pi$, but also the Monte Carlo approximation error :

$$\frac{1}{\lfloor N\alpha \rfloor} \sum_{j, d_j \leq d_{(\lfloor N\alpha \rfloor)}} \delta_{\widetilde{\theta}_j} \approx \pi_{\alpha,d} \tag{2}$$

needs to be considered. In Eq. (2), $\delta$ denotes the Dirac measure and $d_j$ and $d_{(.)}$ are defined in Algorithm 1. Here it only corresponds to the rate of convergence of the empirical measure in the classical law of large number. This question has recently attracted a great deal of attention [Fournier and Guillin, 2015; Dereich et al., 2013; Weed and Bach, 2019]. There is an unavoidable curse of dimensionality : the empirical measure over a $n$-sample (here $n = \lfloor N\alpha \rfloor$) is at distance $n^{-1/d}$ to the target measure. This explains why the standard ABC lacks efficiency when the dimension of $\mathbf{x}$ is large or when the dimension of the summaries is large. This also induces high computation times. Using a small number of relevant summaries is then crucial, but, when the size of the data tends to infinity, these summaries should converge toward an injective mapping from parameters to summaries. This injectivity depends on both the true structural model and the particular choice of $\eta(\cdot)$, hence there is no general method for finding such statistics $\eta(\cdot)$. This can restrict the (computational) use or the validity of the standard ABC algorithm, because without this condition, posterior concentration is not guaranteed.

## 2.2 Sequential Monte Carlo ABC (ABC-SMC)

The standard ABC can be quite inefficient, as numerous draws may be rejected before accepting one. One of the most widely used alternative, now considered the "gold standard" among ABC methods, is the Sequential Monte Carlo ABC (ABC-SMC, Sisson et al. [2007]; Beaumont et al. [2009]; Del Moral et al. [2012]). In this approach, several weighted populations of samples are generated sequentially, these populations being associated with a decreasing sequence of tolerance levels $\epsilon_1 > \epsilon_2 > \ldots > \epsilon_T$. At each iteration, asample of population $t$ is obtained by randomly drawing a sample from the weighted population at iteration $t-1$ and perturbing it. As the iterations increase and the tolerance levels decrease, these weighted populations approximate the target posterior distribution more and more closely. As in the standard ABC approach, summary statistics are used to summarize the samples. This algorithm is briefly described and discussed in appendix E.1, see Algorithm 3. But it is important to note that compared to standard ABC, this approach is sequential and no longer based on a reference table. As a result, we cannot simulate all the required samples in advance and reuse them for any observation of interest. Another consequence of this sequential accept-reject scheme is that the total number of samples needed to reach a given tolerance level is not known beforehand. Moreover, this approach requires a sequence of decreasing tolerance levels, with the final level defining the approximation of the posterior, or at least an initial level $\epsilon_1$ and a method for decreasing the level between populations.

It also requires a transition kernel to perturb samples randomly drawn from the previous population. The construction of such kernels, also referred to as proposal samplers, has been studied in particular by Filippi et al. [2013].

## 2.3 Towards a preprocessing summary statistics free ABC

In this section, we briefly review two attempts in the ABC literature to obtain methods depending less on the summary statistics choice (and distance) than the original ABC method.

### 2.3.1 ABC-Convolutional Neural Network [Akesson et al., 2022]

The earliest work to bypass the choice of summary statistics in the ABC method is maybe the one of Fearnhead and Prangle [2012]. Indeed, they showed that the optimal summary statistic is the true posterior mean of the parameter of interest (for quadratic losses). Although it cannot be calculated analytically, they proposed an extra stage of simulation to estimate it. This estimate is then used as a summary statistic within a standard ABC algorithm. In practice, they used linear regression to estimate the posterior mean, with appropriate functions of the data as predictors. This approach has inspired Jiang et al. [2017] who proposed to estimate the posterior means using deep Neural Networks (NN) instead of linear regression. Akesson et al. [2022] go a bit further by proposing to estimate them using Convolutional Neural Networks (CNN). Indeed, CNNs are often more effective than classical dense NNs for tasks involving structured grid data, such as images, audio, video and time-series data (see for example LeCun et al. [1999] or Boston et al. [2022]). The approach of Akesson et al. [2022] is then a slight modification of the standard ABC: CNN are used to estimate the posterior mean of a parameter, which is then used as a summary statistic. Hence, to have asymptotic consistency and valid frequentist coverages, the same kind of conditions are needed as for Algorithm 1. For sake of completeness, this algorithm is briefly described and commented in appendix E.2, see Algorithm 4.

### 2.3.2 ABC-Random Forest [Raynal et al., 2018]

Another interesting approach is the one of Raynal et al. [2018]. Their goal is slightly different from the goal of standard ABC, as they do not seek to approximate the full posterior distribution. Instead, they focused on functionals of the posterior distribution, like posterior mean, posterior variance or posterior quantiles. Indeed, they are easier to interpret and report than the whole posterior distribution, and are the main interest of practitioners. Formally, they predict a scalar transform $T(\theta)$ of a multidimensional parameter $\theta$. Without loss of generality, their interest is in

$$\psi(\mathbf{x}) = \mathbb{E}_\pi[T(\theta) \mid \mathbf{x}] = \int_\Theta T(\theta)\pi(\theta \mid \mathbf{x})\mathrm{d}\theta = c(\mathbf{x}) \int_\Theta T(\theta) f(\mathbf{x} \mid \theta)\pi(\theta)\mathrm{d}\theta, \tag{3}$$

where $c(\mathbf{x})$ is a normalizing constant. From this functional, we easily recover posterior mean, posterior variance or posterior quantiles using respectively $T(\theta) = \theta$, $T(\theta) = \theta^2$, $T(\theta) = \theta$, and $T(\theta) = \mathbf{1}_{\theta \leq q}$.

To mitigate the curse of dimensionality, the dimension of (3) can be reduced by using summary statistics, i.e., the focus is now on

$$\psi_\eta(\mathbf{x}) = \mathbb{E}_\pi[T(\theta) \mid \eta(\mathbf{x})] = \int_\Theta T(\theta)\pi\{\theta \mid \eta(\mathbf{x})\}\mathrm{d}\theta = c_\eta(\mathbf{x}) \int_\Theta T(\theta) f\{\eta(\mathbf{x}) \mid \theta\}\pi(\theta)\mathrm{d}\theta, \quad (4)$$

which is expected to be a good approximation of $\psi(\mathbf{x})$ but much simpler to estimate. Although many studies focus on how to calculate efficiently an average from a sample (see for example Gobet et al. [2022] and references therein), the idea of directly approximating the functional of interest in Eq. (3) (or similarly in Eq. (4)) bypasses the costly step of approximating the empirical measure Eq. (2), especially when the dimension is large.

Raynal et al. [2018] propose approximations of $\psi_{\boldsymbol{\eta}}(\mathbf{x})$, of the posterior variance $\mathbb{V}_\pi[T(\theta) \mid \eta(\mathbf{x})]$ and of posterior quantiles, using (regression) random forests [Breiman, 2001]. Their algorithm is briefly described and commented in appendix E.3, Algorithm 5. A drawback of this method (as presented in Raynal et al. [2018] ) is that it only works on unidimensional parameter inference. When the parameter $\theta$ of interest is multidimensional, Raynal et al. [2018] recommend constructing one random forest (RF) for each unidimensional component of $\theta$, and covariance between components is not taken into account. Additional RFs might be constructed if one is interested in estimating posterior covariance between pairs of components. Piccioni et al. [2022] used, for instance, this approach with an additional sensitivity analysis. To quantify the uncertainty of their estimation, Raynal et al. [2018] propose to use their approximated estimation of quantiles and variances (in Raynal et al. [2018, Section 2.3.3]) but also to use the so-called *out-of-bag* simulations inherent from RF algorithms (in Raynal et al. [2018, Section 2.3.4]). This latter gives a first hint to quantify uncertainty without giving nevertheless precise confidence intervals.

# 3 The ABCD-Conformal algorithm

In this section, we propose a new ABC method based on Deep learning with Dropout and conformal prediction, entitled ABCD-Conformal.

## 3.1 Motivation and principle

Like Raynal et al. [2018], instead of approximating the whole posterior $\pi(\theta \mid \mathbf{x})$, we are interested in estimating functional of the form $\psi(\mathbf{x}) = \mathbb{E}_\pi[T(\theta) \mid \mathbf{x}]$ (see Eq. (3)). The ABCD-Conformal method will both output a prediction and an associated exact confidence set. Our method uses Neural Networks (NN), similarly to Akesson et al. [2022]. However, in contrast to their approach, the NN directly outputs the functional of interest and an associated uncertainty, enabling us to bypass the nearest neighbors step detailed in Section 2.1 to overcome the shortcomings of standard ABC.

To use conformal prediction, we need both a prediction and an associated measure of uncertainty that we will name *heuristic uncertainty* and denote $\widehat{\mathbb{V}}(\mathbf{x})$. To this end, we will use neural networks with Dropout layers. Dropout layers are classically used to prevent overfitting for neural networks, see Hinton et al. [2012] and Srivastava et al. [2014a]. However, they can also be used to associate uncertainties e.g., variance with predictions from neural networks, see the Bayesian Neural Network literature [Gal, 2016]. We can then obtain valid confidence sets using these uncertainties through conformal prediction [Angelopoulos and Bates, 2023]. These two steps are described in Sections 3.2 and 3.3.

## 3.2 Prediction and heuristic uncertainty using Monte Carlo Dropout

For an introduction to Deep Learning, we refer the reader to Courville et al. [2016]; Guéron [2019] and references therein. Neural networks have a strong predictive power, however they are often regarded as black boxes making their predictions not explainable and without associated uncertainty. This is an issue in many areas like medicine or autonomous vehicles, where false predictions can have big consequences [Filos et al., 2020]. A lot of efforts have been made recently either to explain predictions, see the Explainable AI [Xu et al., 2019; Dantas et al., 2023] literature or to associate an uncertainty with the prediction, see Izmailov et al. [2021] or Kompa et al. [2021] among many others approaches.

In this paper we focus on the Dropout method proposed in Gal [2016] and Gal and Ghahramani [2016] even if more recent uncertainty methods might be more powerful as for example Variational BNN methods [Folgoc et al., 2021; Jospin et al., 2022]. We choose the classical Dropout method as it is generic, simple to implement and to integrate into an already existing NN architecture. As explained in Section 3.3, using conformal prediction will enable us to compensate the shortcomings of Dropout. At every training step, in each Dropout layer, some neurons are randomly dropped (their corresponding weights are set to zero) according to some user-choice parameters (such as the Dropout probability). This mechanism prevents complex co-adaptations on the training data. Co-adaptation means that different neurons have highly correlated behavior. Dropout avoid this, because a neuron cannot rely on other neurons being present, hence neurons are encouraged to detect features independently of each other. This can also be seen as a regularization method during the training, to prevent overfitting. Standard Dropout layers are deactivated during test/prediction phase. Details concerning dropout procedures can be found in [Hinton et al., 2012; Srivastava et al., 2014b]. In their seminal paper Gal and Ghahramani [2016] proposed to activate Dropout during test, introducing randomness in the NN, hence in the output. Repeating the same prediction task several times produces a distribution and is sometimes called Monte Carlo Dropout (MC Dropout).

We briefly describe the MC Dropout procedure. Let $D = \{(\mathbf{x}_j, \theta_j), j = 1, \ldots, N\}$ be the training set, where $\mathbf{x}_j$ corresponds in the ABC framework to the sampled data and $\theta_j$ (or a transformation $T(\theta_j)$) to the unknown parameter of interest. The vector of parameters of the network (weights and bias) is denoted $\omega$. During training, the goal is to find parameters of the network $\omega$ that are likely to have generated the outputs $(\theta_j, j = 1, \ldots, N)$, given the inputs $(\mathbf{x}_j, j = 1, \ldots, N)$. For this goal, we use another Bayesian inference model to infer such $\omega$ and we then aim to estimate the distribution $\pi(\omega \mid D)$. After training, the goal is then to predict the parameter of interest $\theta$ associated to a new $\mathbf{x}$.

**The posterior distribution of interest** is then:

$$\pi(\theta \mid \mathbf{x}, D) = \int \pi(\theta \mid \mathbf{x}, \omega)\pi(\omega \mid D)\,\mathrm{d}\omega. \tag{5}$$

The density $\pi(\omega \mid D)$ is quite complex and cannot be evaluated analytically. It is approximated using a variational approach by $q(\omega)$. **The approximate posterior distribution of interest** is then given by

$$q(\theta \mid \mathbf{x}) = \int \pi(\theta \mid \mathbf{x}, \omega)q(\omega)\,\mathrm{d}\omega. \tag{6}$$

The first two moments of this distribution can be estimated empirically following Monte Carlo integration with $K$ i.i.d. samples with law $q(\omega)\mathrm{d}\omega$. For each Monte Carlo sample $k$, different units of the network are randomly dropped out [Hinton et al., 2012]; we note $\widehat{\omega_k}$ the estimated weights of the associated network with dropped units.Hence, these weights $\widehat{\omega_k}$ are different for each Monte Carlo iteration. Denote by $\mathbf{f}^{\omega}(\mathbf{x})$ the neural network's stochastic output for input $\mathbf{x}$ and parameters $\omega$. If we assume that $\theta \mid \mathbf{x}, \omega \sim \mathcal{N}(\mathbf{f}^{\omega}(\mathbf{x}), \tau^{-1}\mathbf{I})$ and $\widehat{\omega_k} \sim q(\omega)$, $k = 1, \ldots, K$, an unbiased (consistent) estimator for $\mathbb{E}_{q(\theta|\mathbf{x})}[\theta \mid \mathbf{x}]$ is given by:

$$\widehat{\theta}(\mathbf{x}) = \frac{1}{K}\sum_{k=1}^{K} \mathbf{f}^{\widehat{\omega_k}}(\mathbf{x}), \tag{7}$$

which corresponds to the average of $K$ stochastic forward passes through the network with Dropout. We refer the reader to the proposition 2 on page 47 of Yarin Gal's thesis [Gal, 2016] for more details about how to obtain this expression and the followings.
An unbiased (consistent) estimator for the second moment $\mathbb{E}_{q(\theta|\mathbf{x})}[\theta^T\theta]$ is given by:

$$\widehat{\mathbb{E}}[\theta^T\theta \mid \mathbf{x}] = \tau^{-1}\mathbf{I} + \frac{1}{K}\sum_{k=1}^{K} \mathbf{f}^{\widehat{\omega_k}}(\mathbf{x})^T\mathbf{f}^{\widehat{\omega_k}}(\mathbf{x})$$

Then to obtain a variance associated with the prediction of $\theta$ we can use the following unbiased (consistent) estimator:

$$\widehat{\mathbb{V}}[\theta \mid \mathbf{x}] = \underbrace{\tau^{-1}\mathbf{I}}_{\widehat{\mathbb{V}}_a[\theta|\mathbf{x}]} + \underbrace{\frac{1}{K}\sum_{k=1}^{K} \mathbf{f}^{\widehat{\omega_k}}(\mathbf{x})^T\mathbf{f}^{\widehat{\omega_k}}(\mathbf{x}) - \widehat{\theta}(\mathbf{x})^T\widehat{\theta}(\mathbf{x})}_{\widehat{\mathbb{V}}_e[\theta|\mathbf{x}]}, \tag{8}$$

which corresponds to the inverse model precision plus the sample variance of $K$ stochastic forward passes through the network with Dropout. Hence, using Dropout and its associated randomness, we can obtain an estimate of $\mathbb{E}_{q(\theta|\mathbf{x})}[\theta]$, as well as associated uncertainty. This estimate can be considered as an approximation of $\mathbb{E}_{\pi}[\theta]$ because $q(\theta \mid \mathbf{x})$ is an approximation of the posterior of interest $\pi(\theta \mid \mathbf{x}, D)$. Moreover, Gal [2016] interprets the uncertainty Eq. (8) as the sum of an aleatoric uncertainty $\widehat{\mathbb{V}}_a[\theta \mid \mathbf{x}]$ and an epistemic uncertainty $\widehat{\mathbb{V}}_e[\theta \mid \mathbf{x}]$. The aleatoric uncertainty $\widehat{\mathbb{V}}_a[\theta \mid \mathbf{x}]$ is interpreted as the noise in the data, it is the result of measurement imprecision (it is often modelled as part of the likelihood, and this is often a Gaussian corrupting noise). The epistemic uncertainty $\widehat{\mathbb{V}}_e[\theta \mid \mathbf{x}]$ is interpreted as the model uncertainty: uncertainty in the model parameters or in the structure of the model. The total predictive uncertainty $\widehat{\mathbb{V}}[\theta \mid \mathbf{x}]$ combines these both types of uncertainties and will be used as our heuristic uncertainty $\widehat{\mathbb{V}}(\mathbf{x})$. In practice, for each of the $K$ Monte Carlo iterations, the weights $\widehat{\omega_k}$ are different, and the NN with input $\mathbf{x}$ gives as outputs $\mathbf{f}^{\widehat{\omega_k}}(\mathbf{x})$ and $\tau^{-1}_{\widehat{\omega_k}}$. The aleatoric uncertainty $\tau^{-1}$ is estimated by the mean of the $\tau^{-1}_{\widehat{\omega_k}}$, and the epistemic uncertainty by the sample variance of the $\mathbf{f}^{\widehat{\omega_k}}(\mathbf{x})$.

The main drawback of Dropout layers is that the Dropout probability is a new model hyperparameter. To circumvent this issue, Gal et al. [2017] propose a Dropout variant, called Concrete Dropout, that allows for automatic tuning of this probability in large models, using gradient methods. It improves performance and tunes automatically the Dropout rate, producing better calibrated uncertainties

**Remark 1.** *Gal [2016] and Gal and Ghahramani [2016] showed that optimizing any neural network with Dropout is equivalent to a form of approximate inference in a probabilistic interpretation of the model. In other words, the optimal parameters found through the optimization of a Dropout neural network are the same as the optimal variational parameters in a probabilistic Bayesian neural network with the same structure (the parameters are found through a variational inference approach). It means that a network trained with Dropout is equivalent to a Bayesian Neural Network and possesses all the properties of such a network (see MacKay [1992] or Neal [2012] concerning Bayesian Neural Networks). For this equivalence to be true, only one condition should be verified in the variational inference approach, which is about the Kullback-Leibler divergence between the prior distribution of the parameters $\omega$ and an approximating distribution for $\omega$. We will not need to verify such assumption because the theoretical guaranty of our confidence sets will be assured through the conformal prediction.*

## 3.3 Conformal prediction

For a test input $\mathbf{x}$, the NN with concrete Dropout method enables us to obtain an approximation $\widehat{\theta}(\mathbf{x})$ of $\mathbb{E}_\pi[\theta \mid \mathbf{x}]$, as well as associated heuristic uncertainties $\widehat{\mathbb{V}}(\mathbf{x})$. However, it does not give valid confidence sets. In order to obtain such confidence sets, we propose to apply a conformal procedure, as explained in Angelopoulos and Bates [2023]. Conformal prediction is indeed a straightforward way to generate valid confidence sets for a wide variety of models from a proxy of the uncertainty. The better this proxy will be, the smaller the confidence interval will be. This procedure requires a level $\delta \in [0, 1]$ and a small amount of additional calibration data $\big((\theta_j, \mathbf{x}_j), j = 1, \ldots, N_{\text{cal}}\big)$, hundreds of calibration data are theoretically enough (see Eq. (12) below). We propose to use as heuristic uncertainty $\widehat{\mathbb{V}}(\mathbf{x})$ one of the variances obtained from the NN with Dropout in Section 3.2, see Eq. (8). This heuristic uncertainty can be transformed into a rigorous confidence interval $\mathcal{C}(\mathbf{x})$ (with confidence level $1 - \delta$) through the following conformal procedure.

Using the heuristic uncertainty $\widehat{\mathbb{V}}(\mathbf{x})$, we compute a score $s(\theta, \mathbf{x})$ for each data sample in the calibration set, from these scores we can define $\widehat{q}$ the conformal quantile and then a confidence set. More precisely, as in Messoudi et al. [2022], the score function will be

$$s(\theta, \mathbf{x}) = \sqrt{(\theta - \widehat{\theta}(\mathbf{x}))^t \widehat{\mathbb{V}}(\mathbf{x})^{-1} (\theta - \widehat{\theta}(\mathbf{x}))}. \tag{9}$$

We can then define $\widehat{q}$ as the $\frac{\lceil (N_{\text{cal}}+1)(1-\delta) \rceil}{N_{\text{cal}}}$ quantile of $\left\{ s(\theta_j, \mathbf{x}_j), j = 1, \ldots, N_{\text{cal}} \right\}$ to have as the confidence set :

$$\mathcal{C}(\mathbf{x}) = \{\theta \mid s(\theta, \mathbf{x}) \leq \widehat{q}\}. \tag{10}$$

This corresponds to an ellipsoid whose center is $\widehat{\theta}(\mathbf{x})$ and whose covariance matrix is $\widehat{\mathbb{V}}(\mathbf{x})^{-1}/\widehat{q}^2$. In one dimension, we have $s(\theta, \mathbf{x}) = |\theta - \widehat{\theta}(\mathbf{x})| / \sqrt{\widehat{\mathbb{V}}(\mathbf{x})}$, and

$$\mathcal{C}(\mathbf{x}) = \left[ \widehat{\theta}(\mathbf{x}) - \widehat{q}\sqrt{\widehat{\mathbb{V}}(\mathbf{x})}; \widehat{\theta}(\mathbf{x}) + \widehat{q}\sqrt{\widehat{\mathbb{V}}(\mathbf{x})} \right]. \tag{11}$$

This confidence set satisfies the marginal coverage property :

$$1 - \delta \leq \mathbb{P}[\theta \in \mathcal{C}(\mathbf{x})] \leq 1 - \delta + \frac{1}{N_{\text{cal}} + 1}, \tag{12}$$

as shown in Angelopoulos and Bates [2023, Appendix D] (in particular the right-hand side inequality needs that random variables $s(\theta_j, \mathbf{x}_j)$ to be continuous). This probability is marginal over the randomness in the calibration and test points [Angelopoulos and Bates, 2023], see Algorithm 2. One great advantage of this procedure is that it is easy to implement, fast and generic. Moreover, the obtained prediction intervals are non-asymptotic, with no distributional or model assumptions. The only assumption needed is that the calibration data and the test data are i.i.d [Vovk et al., 1999] which is naturally the case in our setting. Finally, this procedure provides a frequentist perspective on a Bayesian approach, reconciling the two points of view.
The performance of this approach in terms of intervals length depends only on the quality of the uncertainty measure used. A "good" uncertainty measure should reflect the magnitude of model error: it should be smaller for easy input, and larger for hard ones. Hard inputs here can refer to extreme points in the training set e.g. see example in section 4.2 where some $\theta_3$ can be very large, and/or with high aleatoric uncertainty. If the uncertainty measure is "bad", then the intervals obtained by the conformal procedure will be quite large and their lengths will not differ a lot between easy and hard inputs. From the previous MC Dropout method, three types of uncertainties were obtained: epistemic, aleatoric and overall (sum of aleatoric and epistemic).

**Remark 2.** *Note that a conditional coverage would be preferable, that is:*

$$\mathbb{P}[\theta \in \mathcal{C}(\mathbf{x}) \mid \mathbf{x}] \geq 1 - \delta. \tag{13}$$

*Indeed, marginal coverage does not guarantee the coverage of $\mathcal{C}(\mathbf{x})$ for any specific $\mathbf{x}$, only the average coverage over the whole domain. However, this conditional property is not guaranteed by the conformal procedure in general, but some metrics can be used to check how close we are to this stronger property, see Angelopoulos and Bates [2023].*

### 3.4 Implementation of ABCD-Conformal

The training step of the proposed method is as follows. a) Generate the training dataset i.e., reference table in ABC vocabulary and the calibration dataset, of sizes $N_{\text{train}}$ and $N_{\text{cal}}$ respectively. b) The NN with Dropout is trained using the reference table. In practice the aleatoric uncertainty $\widehat{\mathbb{V}}_a$ is learned while training the neural network with an heteroscedastic loss as in [Kendall and Gal, 2017, section 3]. One could select the network architecture using an extra validation dataset. We refer the reader to the many tutorials and books available on the implementation of neural networks. c) Monte Carlo Dropout prediction is performed, that is, each sample of the calibration dataset is passed through the trained network with Dropout, $K$ times. For each of these samples, an approximation of $\mathbb{E}_\pi[\theta \mid \mathbf{x}]$ is obtained as the average of the $K$ predictions (see Eq. (7)), associated with an *heuristic uncertainty* which is here the variance, see Eq. (8). d) Using this approximation and associated uncertainty, the score of each sample of the calibration dataset is calculated. The conformal quantile is then the $\frac{\lceil (N_{\text{cal}}+1)(1-\delta)\rceil}{N_{\text{cal}}}$ quantile of the calibration scores. Then, for a new data sample $\mathbf{x}$, to approximate $\mathbb{E}_\pi[\theta \mid \mathbf{x}]$ and obtain an associated confidence interval: e) Monte Carlo Dropout prediction is performed to obtain an approximation of $\mathbb{E}_\pi[\theta \mid \mathbf{x}]$ (average of the predictions) and the associated heuristic uncertainty (variance of the predictions). A confidence interval is then calculated using the conformal quantile calculated in d) Eq. (11).

The pseudocode of the algorithm proposed is detailed in Algorithm 2.

## 4 Applications

In this section, we will illustrate our method on three examples: The Moving Average 2 (MA2) model (Section 4.1), a Lotka-Volterra model (Section 4.2) and a model for lake ecosystem dynamics (Section 4.3). An additional example about Gaussian random field is given in the appendix D. For all these examples, we compare the performances of ABCD-Conformal with those of standard ABC, ABC-SMC, ABC-CNN (Akesson et al. [2022]) and ABC-RF (Raynal et al. [2018]) when possible.

**Datasets.** For each example, we simulate a reference table (training set) of size $N_{\text{train}}$. For ABC-CNN and ABCD-Conformal we need a validation set of size $N_{\text{val}}$, plus a calibration set of size $N_{\text{cal}}$ for ABCD-Conformal. We will test these methods on a test set of size $N_{\text{test}}$. The sizes used in the different examples are given in Table 1). These datasets are the same for the different methods studied. To approximate the posterior, standard ABC and ABC-CNN are used with an $\alpha$ acceptance ratio, hence $\alpha \times N_{\text{train}}$ generated samples are kept to approximate the posterior distribution. Concerning ABC-SMC, the approach is not based on a reference table, hence we can not use the same reference table as for the other methods. Moreover, several choices must be made to implement the algorithm, such as how to decrease the sequence of tolerance levels, or how to select the transition kernel used to perturb a random draw from a previous population. Our choices can be found at the end in Appendix E.1.

**Algorithm 2:** Pseudocode for ABCD-Conformal

**Input :** A Bayesian parametric model $\{f(\cdot \mid \theta), \pi\}$, a data sample $\mathbf{x}$, integers $N_{\text{train}}$, $N_{\text{cal}}$ and $K$ representing sizes of training and calibration sets, and the number of stochastic passes for the Dropout. $\delta$ between 0 and 1 to obtain $(1-\delta)$ confidence sets.

**Output:** An approximation of the posterior expected value $\mathbb{E}_\pi[\theta \mid \mathbf{x}]$ and a confidence interval for $\theta$.

1 **a) Generation of a reference table (training dataset) and a calibration dataset :**

2 For the reference table for instance:

3 **for** $j \leftarrow 1$ **to** $N_{\text{train}}$ **do**

4 | Draw $\theta_j \sim \pi$;

5 | Draw synthetic sample $\mathbf{x}_j = (x_{1,j}, \ldots, x_{d,j})^\top$ from the model $f(\cdot \mid \theta_j)$;

6 **end**

7 **b) Train a Neural Network with concrete Dropout on the reference table**: the training pairs are the $\{(\mathbf{x}_j, \theta_j), j = 1, \ldots, N_{\text{train}}\}$ and the loss is the heteroscedastic loss as in [Kendall and Gal, 2017, Eq. (8)].

8 **c) Monte Carlo Dropout prediction on the calibration set**:

9 **for** $j \leftarrow 1$ **to** $N_{\text{cal}}$ **do**

10 | **for** $k \leftarrow 1$ **to** $K$ **do**

11 | | $\mathbf{x}_j$ is given as input to the trained network to obtain outputs $\mathbf{f}^{\widehat{\omega_k}}(\mathbf{x}_j)$ and $\tau_{\widehat{\omega_k}}^{-1}$;

12 | **end**

13 | Obtain $\widehat{\theta}_j$ and $\tau^{-1}$ by averaging the $\mathbf{f}^{\widehat{\omega_k}}(\mathbf{x}_j)$ outputs (see eq. (7))) and the $\tau_{\widehat{\omega_k}}^{-1}$ ;

14 | and an associated uncertainty $\widehat{\mathbb{V}}(\mathbf{x}_j)$ that can be $\widehat{\mathbb{V}}[\theta_j \mid \mathbf{x}_j]$ (see eq. (8)).

15 **end**

16 **d) Computation of the conformal quantile on the calibration set :**

17 **for** $j \leftarrow 1$ **to** $N_{\text{cal}}$ **do**

18 | Compute the calibration score $s_j$ using the score function, see eq. (9)

$$s_j = \sqrt{(\theta_j - \widehat{\theta}_j)^t \widehat{\mathbb{V}}(\mathbf{x}_j)^{-1} (\theta_j - \widehat{\theta}_j)},$$

19 **end**

20 Compute the conformal quantile $\widehat{q}$ as the $\frac{\lceil (N_{\text{cal}}+1)(1-\delta) \rceil}{N_{\text{cal}}}$ quantile of the calibration scores $s_1, \ldots, s_{N_{\text{cal}}}$.

21 **e) For the new data sample x, approximation of $\mathbb{E}_\pi[\theta \mid \mathbf{x}]$ and confidence set for $\theta$ :**

22 **for** $k \leftarrow 1$ **to** $K$ **do**

23 | $\mathbf{x}$ is given as input to the trained network, to obtain an output $\mathbf{f}^{\widehat{\omega_k}}(\mathbf{x})$;

24 **end**

25 Obtain $\widehat{\theta}(\mathbf{x})$ an approximation of $\mathbb{E}_\pi[\theta \mid \mathbf{x}]$ by averaging these outputs (see Eq. (7)), and an associated uncertainty $\widehat{\mathbb{V}}(\mathbf{x})$ (using Eq. (8));

26 The confidence set for $\theta$ is an ellipsoid whose center is $\widehat{\theta}(\mathbf{x})$ and covariance matrix is $\widehat{\mathbb{V}}(\mathbf{x})^{-1}/\widehat{q}^2$, see Eq. (10).

**Evaluation.** The type of results obtained with the different methods are not exactly the same. Using the standard ABC approach, the ABC-SMC and the ABC-CNN, we obtain approximations of the whole posterior distributions. Using ABC-RF or ABCD-Conformal approaches, we obtain directly estimates of some posterior functional estimate of the parameter of interest $\theta$, in particular the posterior expected value $\mathbb{E}_\pi[\theta \mid \mathbf{x}]$ which can be used to estimate the true parameter $\theta$. To compare the methods, we then compute an estimate of the posterior expected value from the approximated posterior distribution obtained with standard ABC, ABC-CNN and ABC-SMC. Performing that for the $N_{\text{test}}$ datasets of the test set, we can compare the obtained estimates with the true values of the parameters used to generate these datasets. We compute NMAE as well as standard deviations of the absolute differences between the estimates and the true values of the parameters. More precisely, for a parameter of interest $\theta$, we compute

$$\text{NMAE} = \frac{\sum_{i=1}^{N_{\text{test}}} \mid \theta_i - \widehat{\theta}_i \mid}{\sum_{i=1}^{N_{\text{test}}} |\theta_i|},$$

where $\theta_i$ is the true value of $\theta$ for the $i^{th}$ dataset, and $\widehat{\theta}_i$ the estimate of $\theta$ for the $i^{th}$ dataset.

Let us give few comments on this error estimation. First, since the ABC method aims to provide a posterior mean estimator, it may seem odd to compare its output with the true parameter value rather than with the posterior mean, as these two quantities can differ substantially. However, in the (non-trivial) examples we consider, the exact posterior means are unknown and cannot be used as a reference for comparison. Moreover, the likelihoods associated to our examples are based on a large number of observations ($p = 100$ for the MA2 model, a grid of size $100 \times 100$ for the Gaussian random field, $36 \times 2$ observation points for the Lotka-Volterra model and 365 observation points for the Lake model), and so results such as the Bernstein-von Mises Theorem [Van der Vaart, 1998], or its generalizations, ensure that in our examples the true parameter value and the posterior mean are indeed close. Note, however, that for the MA2 example only, the exact posterior is explicit (although costly to evaluate). We therefore numerically compared the posterior mean with the true value of the parameter in our Julia Notebook (see supplementary materials) to confirm this claim. Figure A.2 in Appendix A illustrates this comparison for the 1000 samples in the test set.

Note that this NMAE is defined for unidimensional parameters. Hence, in case of parameters of interest that are multidimensional, we calculate it for each marginal of $\theta$ separately. For standard ABC and ABC-CNN, $\widehat{\theta}_i$ corresponds to the empirical mean of the $\theta$ corresponding to the $\alpha N_{\text{train}}$ samples kept to approximate the posterior distribution of $\theta$ given the $i^{th}$ dataset. For ABC-SMC, it corresponds to the weighted mean of the last population. For ABCD-Conformal, $\widehat{\theta}_i$ is the empirical mean of $N_{\text{val}}$ stochastic forward passes through the network with Dropout, given the $i^{th}$ dataset as input (see Eq. (7)). To have an idea of the variability of the absolute errors, we compute $sd(|\theta - \widehat{\theta}|)$ where

$$sd(|\theta - \widehat{\theta}|) = \sqrt{\frac{1}{N_{\text{test}}} \sum_{i=1}^{N_{\text{test}}} (\theta_i - \widehat{\theta}_i)^2}.$$

**Confidence and credible Intervals/Regions.** Finally, as we are interested by the uncertainty associated with an estimate, we compare credible and confidence sets for the five methods. First, we compare confidence and credible sets for each component of $\theta$, hence in dimension one. For standard ABC and ABC-CNN, we used the $\delta/2$ and $1-\delta/2$ quantiles of the marginals of the approximated posterior distributions to obtain credible intervals, for $\delta = 0.05$. For ABC-SMC we did the same thing, but using weighted quantiles. ABC-RF directly gives confidence intervals, for unidimensional parameters. Concerning ABCD-Conformal, we computed confidence intervals for each component of $\theta$, by using the diagonal terms of the uncertainty measure $\widehat{\mathbb{V}}(\mathbf{x})$ (by analogy with a variance-covariance matrix, this corresponds to using only the variance terms). Then, we compare confidence and credible sets in multidimensional terms. Confidence sets (ellipsoids) are directly obtained from the ABCD-Conformal approach. For ABC methods approximating the full multidimensional posterior with neighboring samples, i.e. ABC-Standard, ABC-SMC and ABC-CNN, one could define a 95% credible set using the Highest Posterior Density Region [Hyndman, 1996]. Here for simplicity and computational tractability, we assume the posterior to be a Multivariate Normal distribution. The associated credible set is the ellipsoid generated by Eqs. (9, 10) where $\hat{\mathbb{V}}(\mathbf{x})$ is replaced by the covariance of the neighboring samples and $\hat{q}$ is replaced by the 0.95 quantile of a Chi distribution whose degrees of freedom are the dimension of $\theta$.

The default uncertainty measure used by ABCD-conformal in all the examples is the overall uncertainty returned by the Dropout procedure. For comparison, we also present results using the epistemic uncertainty for the Gaussian random field example (Appendix D), and results using a constant uncertainty for the Lotka Voltera example (Section 4.2).

For all methods, the coverage is estimated by the number of true values of $\theta$ or marginals of $\theta$ in the obtained sets, for the $N_{\text{test}}$ simulations from the test set. We then compare the coverage of these credible and confidence sets for these methods, the mean and median lengths of credible and confidence intervals for marginals of $\theta$, and volumes for multidimensional credible and confidence sets when possible.

**Code and Software** Source codes and instructions to run the examples, in R and/or Julia, are available as supplementary materials. For some examples, datasets and pre-trained models are provided to reproduce exactly our results. For more details, see the dedicated section at the end of the article. For this paper, the authors developed a Julia package `ABCMethods.jl` [Métivier, 2025] where the ABC-Standard, ABC-SMC, ABC-CNN and ABC-Conformal methods and associated tools are coded. All the parameters, methods and languages used in the different examples are synthesized in Table 1.

## 4.1 The Moving Average 2 model

### 4.1.1 The model

The Moving Average 2 model (MA2) is a simple benchmark example used in Bayesian and ABC literature (see Marin and Robert [2007], Marin et al. [2012], Jiang et al. [2017], Wiqvist et al. [2019] or Akesson et al. [2022] for instance). It is defined for observations $X_1, \ldots, X_p$;

| Example | $N_{\text{train}}$ | $N_{\text{val}}$ | $N_{\text{cal}}$ | $N_{\text{test}}$ | $\alpha$ | ABC methods | | | | | Language |
|---|---|---|---|---|---|---|---|---|---|---|---|
| | | | | | | stand | SMC | CNN | RF | D-conf | |
| Section 4.1 MA2 | $10^5$ | $10^3$ | $10^3$ | $10^3$ | 0.001 | ✓ | ✓ | ✓ | ✗ | ✓ | Julia |
| Section 4.2 Lotka-Volterra | $10^5$ | $10^3$ | $10^3$ | $10^3$ | 0.001 | ✓ | ✓ | ✓ | ✗ | ✓ | Julia |
| Section 4.3 Lake model | $3 \times 10^4$ | $10^3$ | $10^3$ | $10^3$ | 0.001 | ✓ | ✓ | ✓ | ✗ | ✓ | Julia |
| Appendix D Gaussian field | $7 \times 10^3$ | 1900 | $10^3$ | $10^2$ | 0.01 | ✓ | ✗ | ✓ | ✓ | ✓ | R |

Table 1: Sizes of datasets used for the different methods: $N_{\text{train}}$ for the training set, $N_{\text{val}}$ for the validation set, $N_{\text{cal}}$ for the calibration set and $N_{\text{test}}$ for the test set, and $\alpha$ the tolerance threshold used for the ABC standard and ABC-CNN. Methods used on the examples: stand. denotes the standard vanilla ABC, SMC the ABC-SMC, CNN the ABC-CNN and D-conf. our ABCD-conformal method. The language column is to specify which language has been used to obtain the presented results, even if code has been developped in both R and Julia.

$$X_j = Z_j + \theta_1 Z_{j-1} + \theta_2 Z_{j-2}, j = 1, \ldots, p,$$

where $(Z_j)_{-2<j\leq p}$ is an i.i.d. sequence of standard Gaussian $\mathcal{N}(0, 1)$. This model is identifiable in the following triangular region $\mathcal{D}$:

$$-2 < \theta_1 < 2, \quad \theta_1 + \theta_2 > -1, \quad \theta_1 - \theta_2 < 1;$$

See for instance Marin and Robert [2007][Chapter 5] or [Marin et al., 2012, Section 2]. We assume that the prior distribution on $(\theta_1, \theta_2)$ is the uniform distribution over this triangular region. This model allows for exact calculation of the posterior $(\theta_1, \theta_2) \mid X$. The goal is to obtain posterior estimates of the bi-dimensional vector of parameters $(\theta_1, \theta_2)$ from an observed dataset of length $p$, $(X_1, \ldots, X_p)$. In the following, we will use $p = 100$.

#### 4.1.2 Algorithm parametrization

Number of used samples are given in Table 1. For the standard ABC, the summary statistics chosen are the first two autocovariances:

$$\tau_1 = \frac{1}{p-1} \sum_{j=2}^{p} x_j x_{j-1} - \left( \frac{1}{p} \sum_{j=1}^{p} x_j \right)^2 \quad \text{and} \quad \tau_2 = \frac{1}{p-2} \sum_{j=3}^{p} x_j x_{j-2} - \left( \frac{1}{p} \sum_{j=1}^{p} x_j \right)^2.$$

To compare two samples, we then use the $\mathbb{L}^2$ distance between the two associated vectors of summary statistics. For ABC-SMC, the same summary statistics and distance are used, the algorithm consists of 10 iterations, in each iteration $10^4$ simulations are made and the best $10^3$ are kept to be the population of this iteration. Hence a total of $10^5$ simulations is made for each test sample, to be comparable with standard ABC and ABC-CNN.

In ABC-CNN method, the distance function between a sample from the test set and a sample from the training set is the quadratic $\mathbb{L}^2$ distance between the parameters predicted by the CNN for the test sample, and true parameters used to simulate the training sample.

Regarding ABC-CNN and ABCD-Conformal, architectures of neural networks are the same for the two methods: 3 convolutional 1D layers with 128 neurons and a kernel size of 2 followed by max-pooling for the first two layers and by a flatten layer for the third one. Then, 3 dense layers of 100 neurons. Each of the previous layers uses the `tanh` activation function. The last layer to obtain the output of dimension 2 is linear. The raw samples of length $p = 100$ are the inputs of the neural networks, and the outputs are the bi-dimensional associated vectors of parameters $(\theta_1, \theta_2)$.

#### 4.1.3 Results

All indicators detailed in the beginning of Section 4 are presented in Table 2. To compare ABCD-Conformal with the other methods, we computed confidence intervals for each component of $\theta$. Figure 1 shows the predicted parameters $\widehat{\theta}_1$ and $\widehat{\theta}_2$ estimated by the ABCD-Conformal method against the true values for the 1000 test samples, and confidence intervals in gray. For illustration, see Figure A.1 in Appendix A, which compares confidence ellipses on a single test sample. On the 1000 test samples the multidimensional coverage obtained using ABCD-conformal is of 94.6%.

| | Standard ABC | ABC-SMC | ABC-CNN | ABC-Conformal |
|---|---|---|---|---|
| NMAE($\theta_1$) | 0.174 | 0.176 | 0.177 | **0.166** |
| $sd(\lvert\theta_1 - \widehat{\theta}_1\rvert)$ | 0.087 | 0.089 | 0.091 | **0.083** |
| mean length $CI(\theta_1)$ | 0.544 | 0.626 | **0.377** | 0.560 |
| median length $CI(\theta_1)$ | 0.535 | 0.609 | **0.375** | 0.559 |
| coverage $CI(\theta_1)$ | 94.0% | **96.5%** | 81.2% | 95.7% |
| NMAE($\theta_2$) | 0.273 | 0.272 | 0.264 | **0.234** |
| $sd(\lvert\theta_2 - \widehat{\theta}_2\rvert)$ | 0.101 | 0.101 | 0.099 | **0.089** |
| mean length $CI(\theta_2)$ | 0.615 | 0.667 | **0.391** | 0.583 |
| median length $CI(\theta_2)$ | 0.596 | 0.653 | **0.394** | 0.585 |
| coverage $CI(\theta_2)$ | 93.2% | 95.8% | 78.1% | **94.8%** |
| mean area $CE(\theta)$ (2D) | 0.437 | 0.507 | **0.187** | 0.409 |
| median area $CE(\theta)$ (2D) | 0.412 | 0.479 | **0.187** | 0.411 |
| coverage $CE(\theta)$ (2D) | **95.1%** | 96.7% | 75.6% | 94.5% |

Table 2: Moving Average 2 example: for each method and each component of $\theta$, we give the NMAE, $sd(\lvert\theta - \widehat{\theta}\rvert)$, the mean and median lengths of the confidence or credible intervals and the coverage. For each method the mean and median areas of multidimensional confidence or credible ellipses are also given, as well as the associated coverage. These indicators are computed on a test set of size $10^3$. The best values for each line are highlighted in bold.

We can see in Table 2 that the four methods obtain quite similar results concerning NMAE and standard deviation of the absolute error, with a slight advantage for ABCD-conformal. Some differences can be noted for the coverages, and the mean lengths of the confidence intervals. The coverage is sharp for standard ABC, ABC-SMC and ABCD-Conformal. However, it is too small for ABC-CNN. This is because the approximated posteriors are too peaked around the mean values, leading confidence intervals with a small coverage (and hence useless). Concerning the mean and median CI lengths, ABCD-conformal and standard ABC give the best results which are quite close (discarding ABC-CNN which does not have good coverage). Concerning the confidence ellipses, ABCD-conformal gives the smallest one (excluding again ABC-CNN). Note that it was expected that standard ABC and ABC-SMC performed quite well, as we used *good* summary statistics. Moreover, as showed by Frazier et al. [2018], conditions for good asymptotic properties are verified for standard ABC on this example, which is quite rare.

To sum up, ABC-CNN does not provide reliable credible intervals, ABCD-Conformal performs slightly better than all other methods in terms of NMAE and coverage (especially for the estimation of $\theta_2$) including the standard ABC and ABC-SMC, which are here in favorable estimation conditions.

**Remark 3.** *We also evaluated the performances of the different methods by comparing the estimated $\theta$ values with the exact posterior means of $\theta$ rather than with the true $\theta$ used to generate the samples. In this example the exact posterior means can indeed be computed. As expected, and as explained at the beginning of section 4, the exact posterior means are very close to the true values, see Figure A.2 in Appendix A. Consequently, the results and performances obtained with the different methods are similar to those presented in Table 2, and the same conclusions can be drawn. See supplementary materials for more details.*

## 4.2 Lotka-Volterra model

### 4.2.1 The model

The Lotka-Volterra model describes the dynamics of biological systems in which two species interact, one as a predator and the other as a prey. Here we consider a stochastic Markov jump process version of this model with state $(X_1, X_2) \in \mathbb{Z}^2$ representing prey and predator population sizes, and three parameters $\theta_1, \theta_2$ and $\theta_3$. Three transitions are possible, with hazard rates $\theta_1 X_1$, $\theta_2 X_1 X_2$ and $\theta_3 X_2$ respectively:

$$
\begin{aligned}
(X_1, X_2) &\xrightarrow{\theta_1, X_1} (X_1 + 1, X_2) \qquad \text{(prey growth)} \\
(X_1, X_2) &\xrightarrow{\theta_2, X_1 X_2} (X_1 - 1, X_2 + 1) \qquad \text{(predation interaction)} \\
(X_1, X_2) &\xrightarrow{\theta_3, X_2} (X_1, X_2 - 1) \qquad \text{(predator death)}
\end{aligned}
$$

The estimation of the parameters of this Lotka-Volterra model has been studied by several authors in an ABC framework, see for instance Prangle [2017]. The initial conditions are $X_1(0) = 50$ and $X_2(0) = 100$ and a dataset corresponds to observations of state $(X_1, X_2)$ at times $0, 2, 4, \ldots, 36$. As usual, all simulations with an extinction of either the preys or the predators were discarded : we are interested in the conditional law of survival. The prior distributions are independent uniforms $\mathcal{U}[-6, 2]$ for the transformed parameters $\log(\theta_1), \log(\theta_2)$ and $\log(\theta_3)$. We are interested here in the estimation of $\theta = (\theta_1, \theta_2, \theta_3)$.

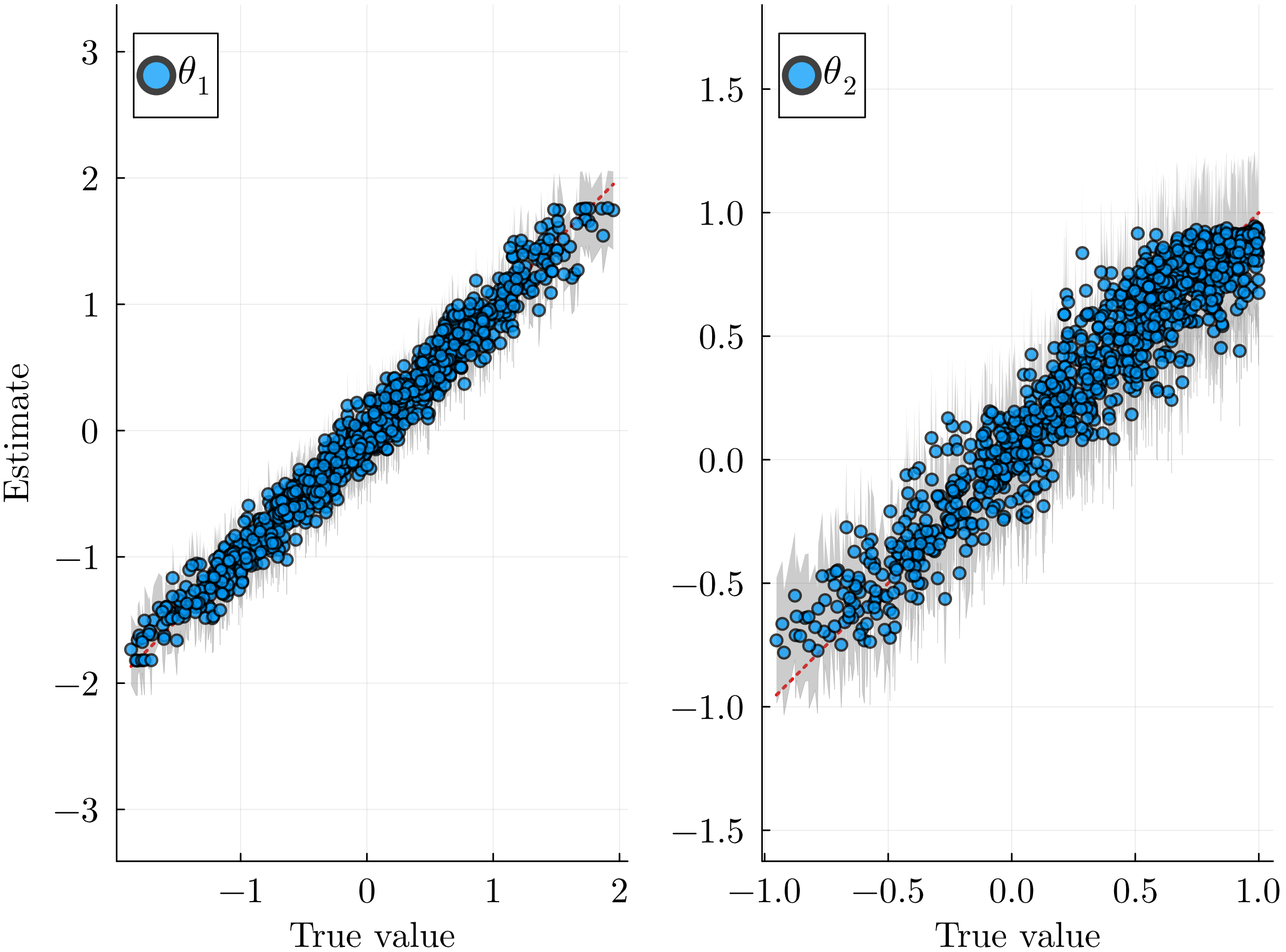

Figure 1: Moving Average 2 example: estimated parameters $\widehat{\theta_1}$ and $\widehat{\theta_2}$ using the ABCD-Conformal against the true values for the 1000 test samples, and confidence sets obtained by the conformal procedure (in gray intervals, for the epistemic uncertainty).

### 4.2.2 Algorithm parametrization

Number of used samples are given in Table 1. The Gillespie's stochastic simulation algorithm is used to generate all the samples. To improve performances of the algorithm, once simulations have been done, we used standardized versions of the parameters. The goal is then to obtain posterior estimates of the tri-dimensional vector of normalized parameters $(\theta_1, \theta_2, \theta_3)$, from an observed sample consisting of two time series of size 19.

Concerning the standard ABC, there is no obvious summary statistic. Instead, the distance function $d\Big((x_1, x_2), (x_{1d}, x_{2d})\Big)$ between a sample from the test set $\big(\{x_{1d}[i], x_{2d}[i]\}, i = 1, \ldots, 19\big)$ and a sample from the training set $\big(\{x_1[i], x_2[i]\}, i = 1, \ldots, 19\big)$ , is given by the sum of squared differences:

$$d\Big((x_1, x_2), (x_{1d}, x_{2d})\Big) = \sum_{i=1}^{19} \Big(\big(x_1[i] - x_{1d}[i]\big)^2 + \big(x_2[i] - x_{2d}[i]\big)^2\Big)$$

For ABC-SMC, the same distance is used, the algorithm consists of 10 iterations, in each iteration $100 \times 100$ simulations are made and the best 100 are kept to be the population of this iteration. Hence a total of $10^5$ simulations is made for each test sample, to be comparable with standard ABC and ABC-CNN. See Appendix E.1 for the implementation details. Concerning ABC-CNN, the distance function between a test sample and a training sample is the quadratic distance between the parameters predicted by the CNN for the test sample, and the true parameters used to simulate the training sample. The architectures used for the neural networks are the same for the ABC-CNN and the ABCD-Conformal: 3 convolutional 1D layers with 128 neurons and a kernel size of 2, followed by max-pooling for the first two layers, and by a flatten layer for the third one. Then, 3 dense layers of 128 neurons. Each of the previous layers uses the `tanh` activation function. The last layer to obtain the output of dimension 3 is linear. The raw samples consisting of two time series of lengths 19 are the inputs of the neural networks, and the outputs are the tri-dimensional associated vectors of parameters $(\theta_1, \theta_2, \theta_3)$. Concerning ABCD-conformal, the default uncertainty measure is the overall uncertainty returned by the Dropout procedure. To show the benefits of using a "relevant" measure of uncertainty, we also present results using a constant uncertainty (only for confidence ellipsoids).

Note that this example is very challenging as $\theta$ is conditionned by the non-extinction condition (some prior values of $\theta$ can lead almost systematically to exctinction). Morevover, $\theta_1$ and $\theta_3$ both have heavy tailed distributions. It means that in some rare cases their values can be very large. In general Machine Learning algorithms are not very robust to these extremes [Drenkow et al., 2022] which can lead to difficulties in the training. In Section 4.2.3, we hence study the results conditional on the value of $\theta$.

### 4.2.3 Results

All indicators detailed in the beginning of Section 4 are presented in Table 3. Looking marginally at each component of $\theta$, we see that overall ABC-SMC is better than standard ABC, on all criteria except coverage, for which the standard ABC is slightly better. We also note that ABC-CNN and ABCD-Conformal outperform standard ABC and ABC-SMC and obtain quite better results than ABC-CNN, when looking at NMAE and standard deviations of the absolute errors. ABCD-Conformal shows similar results to ABC-CNN regarding these indicators, which shows the benefit of using neural networks and no summary statistics to obtain good predictions. Concerning the mean and median lengths of CI, sometimes it is ABC-CNN which is better, other times it is ABCD-conformal. Concerning the coverages of CI, when comparing ABC-CNN and ABCD-conformal we can note that ABC-CNN does not always have a good coverage, unlike ABCD-conformal. Hence, marginally and globally on all three dimensions, ABCD-conformal gives the most acceptable results. Looking at the muldimensional confidence and credible ellipses is quite interesting. Indeed, we can note that ABCD-Conformal give ellipses with the largest volume in mean, but not in median, and that mean and median volumes are very different. ABC-CNN and ABC-SMC are discarded because of their too small coverages.

| | Stand. ABC | ABC-SMC | ABC-CNN | ABCD-Conf. overall | ABCD-conf. constant |
|---|---|---|---|---|---|
| $\theta_1$ parameter | | | | | |
| $NMAE(\theta_1)$ | 0.186 | 0.161 | 0.090 | **0.082** | |
| $sd(\lvert\theta_1 - \widehat{\theta}_1\rvert)$ | 0.503 | 0.527 | 0.276 | **0.260** | |
| mean length $CI(\theta_1)$ | 0.764 | 0.396 | **0.326** | 0.381 | |
| median length $CI(\theta_1)$ | 0.316 | 0.208 | 0.231 | **0.178** | |
| coverage $CI(\theta_1)$ | **96.7%** | 93% | 96.1% | 94.5% | |
| $\theta_2$ parameter | | | | | |
| $NMAE(\theta_2)$ | 0.1882 | 0.1529 | 0.1543 | **0.1403** | |
| $sd(\lvert\theta_2 - \widehat{\theta}_2\rvert)$ | 0.0029 | 0.0029 | **0.0018** | 0.0020 | |
| mean length $CI(\theta_2)$ | 0.0062 | 0.0036 | **0.0028** | 0.0050 | |
| median length $CI(\theta_2)$ | 0.0037 | **0.0024** | **0.0024** | **0.0024** | |
| coverage $CI(\theta_2)$ | **95.6%** | 94.2% | 80.7% | 94.9% | |
| $\theta_3$ parameter | | | | | |
| $NMAE(\theta_3)$ | 0.195 | 0.173 | 0.091 | **0.081** | |
| $sd(\lvert\theta_3 - \widehat{\theta}_3\rvert)$ | 0.438 | 0.461 | **0.218** | 0.222 | |
| mean length $CI(\theta_3)$ | 0.697 | 0.365 | **0.278** | 0.318 | |
| median length $CI(\theta_3)$ | 0.268 | 0.175 | 0.177 | **0.113** | |
| coverage $CI(\theta_3)$ | **97.2%** | 92.4% | 95.7% | 94.6% | |
| $\theta$ parameter (3 dimensional) | | | | | |
| mean volume $CE(\theta)$ | 0.0197 | 0.00269 | **0.00207** | 0.0336 | 0.0230 |
| median volume $CE(\theta)$ | 0.00028 | **0.00008** | 0.0004 | 0.00015 | 0.0230 |
| coverage $CE(\theta)$ | 95.9% | 91.5% | 88.8% | 94.6% | **96.7%** |

Table 3: Lotka-Volterra example: for each method and each component of $\theta$, we give the NMAE, $sd(|c - \widehat{c}|)$, the mean and median lengths of the confidence or credible intervals and the coverage. For each method the mean and median volumes of multidimensional confidence or credible ellipses are also given, as well as the associated coverage. For the ABCD-conformal method, the overall uncertainty is used by default. A constant uncertainty has also been used for confidence ellipses. All the indicators are computed on a test set of size $10^3$. The best values for each line are highlighted in bold.

In this example, to highlight the benefit of using a relevant heuristic uncertainty measure $\hat{\mathbb{V}}(\mathbf{x})$ for the conformal procedure, we also show the results when $\hat{\mathbb{V}}(\mathbf{x})$ is constant, see Table 3. We notice that in this case, the confidence sets also have a good coverage, thanks to the conformal procedure, but they do not adapt to the input as they are all of the same size (hence same mean and median). We therefore see the benefit of using a relevant uncertainty like the overall variance, which reflects the magnitude of model error: the volumes of the credible sets obtained are then small for most (easy) inputs, and large for the few hard ones. This is demonstrated by the median volume of the confidence sets being much smaller than the mean when using the ABC-Conformal method with non constant uncertainty $\hat{\mathbb{V}}(\mathbf{x})$.

As noted before, the distributions of $\theta_1$ and $\theta_3$ have heavy tails. To show that this behaviour yields big differences between mean and median lengths of confidence and credible intervals and ellipses, we investigate the results for different regions of values of $\theta_3$ for illustration. The results are summarized in Table 4, and similar results were obtained for $\theta_1$. We can understand that the performances of the different methods can vary a lot depending on the region of the parameter. For a given component of $\theta$, the coverages on the whole domain are quite different from the coverages on some specific regions. For instance, the coverage of standard ABC (resp. ABCD-Conformal) on the whole domain is 97.2% (resp 94.6%), while it is only 77.5% (resp 87.5%) for extreme parameter region, $\theta_3 > 3$. Hence, even if in theory ABCD-Conformal gives marginal coverage and standard ABC gives conditional coverages, in this example standard ABC also does not give similar coverages depending on the region. Thanks to its good uncertainty heuristic, ABCD-Conformal is able to perform the least bad on harder parameters regions while being very good in easier regions $\theta < 1$ and $1 < \theta < 3$. In these easier regions, it has the best results in terms of size of the confidence sets, compared to standard ABC which is the only other method having a good coverage. Concerning ABC-SMC and ABC-CNN, they should be discarded most of the time because of too small coverages.

To sum up, ABCD-Conformal is most often better than the other methods in this example, with better predictions and smaller confidences intervals or ellipsoids, associated to good coverages. We note that the coverages are quite different in different regions of the parameters, illustrating the marginality of the coverage, but this is the case for all methods.

For illustrative purposes, Figures B.1 and B.2 in Appendix B show estimated $\theta$ associated with 3D-confidence ellipsoids obtained using standard ABC, ABC-CNN and ABCD-conformal, and true values. Figure B.1 shows an example where ABCD-Conformal has the smallest confidence ellipsoid with the true $\theta$ inside. Figure B.2 shows an example where ABC-Conformal has the largest confidence region with the true $\theta$ inside, while other methods fail to capture the difficulty of estimating this parameter and provide too small confidence ellipsoids which do not include the true parameter.

| | Stand. ABC | ABC-SMC | ABC-CNN | ABCD-Conf. overall | ABCD-Conf. constant |
|---|---|---|---|---|---|
| $\theta_3 < 1$, 782 test samples | | | | | |
| mean length $CI(\theta_3)$ | 0.255 | 0.164 | 0.166 | **0.120** | |
| median length $CI(\theta_3)$ | 0.192 | 0.120 | 0.148 | **0.0967** | |
| coverage $CI(\theta_3)$ | 98.1% | 94.4% | **97.1%** | 94.8% | |
| $1 \leq \theta_3 \leq 1$, 178 test samples | | | | | |
| mean length $CI(\theta_3)$ | 1.922 | 0.889 | 0.580 | **0.560** | |
| median length $CI(\theta_3)$ | 1.41 | 0.745 | 0.517 | **0.399** | |
| coverage $CI(\theta_3)$ | 97.8% | 91.0% | 94.4% | 95.5% | |
| $3 < \theta_3$, 40 samples | | | | | |
| mean length $CI(\theta_3)$ | 3.89 | 1.97 | **1.13** | 3.11 | |
| median length $CI(\theta_3)$ | 4.19 | 1.98 | **1.13** | 2.41 | |
| coverage $CI(\theta_3)$ | 77.5% | 60.0% | 75.0% | **87.5%** | |
| $\theta$ parameter (3 dimensional), for $\theta_3 < 1$ | | | | | |
| mean volume $CE(\theta)$ | 0.00124 | 0.00027 | **0.00026** | 0.00074 | 0.0230 |
| median volume $CE(\theta)$ | 0.00013 | **0.00004** | 0.00008 | 0.00007 | 0.0230 |
| coverage $CE(\theta)$ | 97.8% | 93.6% | 91.8% | 95.4% | 99.5% |
| $\theta$ parameter (3 dimensional), for $1 \leq \theta_3 \leq 1$ | | | | | |
| mean volume $CE(\theta)$ | 0.0632 | 0.0077 | **0.0049** | 0.0227 | 0.0230 |
| median volume $CE(\theta)$ | 0.0341 | 0.0020 | **0.0019** | 0.0024 | 0.0230 |
| coverage $CE(\theta)$ | 94.4% | 89.3% | 80.3% | **94.9%** | 94.4% |
| $\theta$ parameter (3 dimensional), for $3 < \theta_3$ | | | | | |
| mean volume $CE(\theta)$ | 0.188 | 0.028 | **0.025** | 0.724 | 0.0230 |
| median volume $CE(\theta)$ | 0.150 | 0.015 | **0.019** | 0.166 | 0.0230 |
| coverage $CE(\theta)$ | 65.0% | 60.0% | 67.5% | 77.5% | 52.5% |

Table 4: Lotka-Volterra example, results conditioned by the value of $\theta_3$: for each method and component $\theta_3$, we give the mean and median lengths of the confidence or credible intervals and the coverage. For each method the mean and median volumes of multidimensional confidence or credible ellipses are also given, as well as the associated coverage. For the ABCD-conformal method, the overall uncertainty is used by default. A constant uncertainty has also been used for confidence ellipses. All the indicators are computed on a test set of size $10^3$. The best values for each line are highlighted in bold.

## 4.3 A model for lake ecosystem dynamics

### 4.3.1 The model

In this section, we address the problem of estimating nine parameters in the system of two differential equations introduced in Piccioni et al. [2022]. Although still a simplified model, it is more complex than those presented in the other sections. It involves a much larger number of parameters whose influence is highly non-linear. In addition, it is a non-dominated model; in particular, this means that it does not admit any likelihood with respect to a reference measure. Consequently, estimating the posterior law for such a model using MCMC methods is completely impossible and a likelihood-free approach, such as ABC method or its variants, is required (see for instance Macci and Spizzichino [2015]). It is worth noting that this difficulty is often overcome by adding random noise to the dynamics, whether or not this is justified for the problem under consideration.

The system under consideration represents a lake ecosystem in which the time evolution of phytoplankton and inorganic phosphorus concentrations is described by the following equations:

$$\frac{\mathrm{d}C_{\mathrm{P}}}{\mathrm{d}t} = \mu(C_{\mathrm{IP}}, T)C_{\mathrm{P}} - d_{\mathrm{P}}C_{\mathrm{P}} \tag{14}$$

$$\frac{\mathrm{d}C_{\mathrm{IP}}}{\mathrm{d}t} = V_{\min}(C_{\mathrm{OP}}, C_{\mathrm{IP}}, T) - a_{\mathrm{pc}}\mu(C_{\mathrm{IP}}, T)C_{\mathrm{P}} + (1 - f_{\mathrm{OP}})a_{\mathrm{pc}}d_{\mathrm{P}}C_{\mathrm{P}}. \tag{15}$$

For the sake of presentation, we do not provide details here about the meaning of parameters; full information is given in the Appendix C. However, it should be noted that $C_{\mathrm{OP}}$ is a linear combination of $C_{\mathrm{P}}$ and $C_{\mathrm{IP}}$, $T$ is a time-periodic function, $\mu$ and $V_{\min}$ are non-linear maps. In this model, there are 15 parameters and 2 initial conditions. We fixed these two initial conditions as well as 6 parameters that are not included in the estimation process. Our objective is to estimate the 9 remaining parameters corresponding to the vector $\theta$ defined as follows: $\theta = (\mu_{\max}, K_{\mathrm{IP}}, d_{\mathrm{P}}, f_{\mathrm{OP}}, V_{\max}, K_{\mathrm{OP}}, K_{\mathrm{I}}, \theta_{\min}, \theta_{\mathrm{gr}})$. A dataset corresponds to observations of state $(C_{\mathrm{P}}, C_{\mathrm{IP}})$ at times $0, 0.5, 1, \ldots, 364.5$ days (i.e. on a one year time-scale, every 12 hours). The prior distributions are independent uniforms.

### 4.3.2 Algorithm parametrization

The Numbers of samples used are given in Table 1. In this model, the different parameters have different scales and units. Therefore, to train the Neural Networks, the training and validation sets were scaled. Concerning standard ABC, there is no obvious choice of summary statistic. Moreover, the trajectories of $C_{\mathrm{P}}(t)$ and $C_{\mathrm{IP}}(t)$ have very different scales. As a consequence, for the distance function between a test sample and a training sample, we chose to use the NMAE to get rid of the scale. For ABC-SMC, the same distance is used. The algorithm consists of 6 iterations, in each iteration $5 \times 10^3$ simulations are performed, and the best $10^2$ are kept to be the population of this iteration. Hence a total of $3 \times 10^4$ simulations is performed for each test sample, in order to ensure comparability with standard ABC and ABC-CNN. Concerning ABC-CNN, the distance function between a test sample and a training sample is defined as the quadratic distance between the parameters predicted by the CNN for the test sample, and the true parameters used to generate the training sample.

Again, the architectures used for the neural networks are similar to the previous applications: 3 convolutional 1D layers with 128 neurons and a kernel size of 2, followed by max-pooling for the first two layers, and by a flatten layer for the third one. Then, we have 3 dense layers of 128 neurons. Each of the previous layers uses the `tanh` activation function. The last layer which leads to the output of dimension 9 is linear. The raw samples consisting of two time series of lengths 730 (for $C_{\mathrm{P}}$ and $C_{\mathrm{IP}}$) are the inputs of the neural network, and the output is the multidimensional associated vector of 9 parameters.

### 4.3.3 Results

All indicators detailed in the beginning of Section 4 are presented in Table 5. In this Table 5, we see that ABCD-Conformal always has a good coverage, while this coverage is sometimes too small for the three other methods.

Looking marginally at each component of $\theta$, when considering NMAE and standard deviations of absolute errors, ABCD-Conformal often outperforms standard ABC, ABC-SMC and ABC-CNN, see for instance $\theta_1$ or $\theta_9$. Concerning the mean and median lengths of the confidence intervals, they are smaller for ABCD-Conformal compared to the other methods, except for $\theta_5$, $\theta_6$ and $\theta_8$. But for these parameters the other methods have too small coverages (except for $\theta_8$ for which ABC-SMC performs particularly well), while the coverage of ABCD-conformal remains good.

Looking at the performances for the 9-dimensional parameter $\theta$, ABCD-conformal gives confidence sets with a quite smaller volume than ABC-standard and ABC-CNN, and with a larger volume than ABC-SMC. However, it is the only method with a satisfactory coverage. Figures C.1, C.2, C.3 and C.4 in appendix C show scatterplots of estimated parameters against the true values for the 1000 test samples and associated confidence sets, for the four methods. This higher-dimensional example clearly illustrates the pitfalls of ABC methods based on summary statistics and threshold selection. Indeed, for standard ABC and ABC-SMC, comparing two large multidimensional time series, even with the informed choice of using the NMAE instead of the naive $\mathbb{L}_2$ norm, is not sufficient to capture the effect of each parameter. Regarding ABC-CNN, which circumvents this issue by using a neural network as summary statistics, it still needs to select neighbors to obtain the posterior distribution. As a result, it either overestimates the size of its confidence intervals compared to ABCD-Conformal or fails to ensure coverage, as seen for $\theta_5, \theta_6$ and $\theta_7$. Both phenomena hold when considering the multidimensional confidence regions.

In such models, some parameters are often not identifiable. In this particular example, the parameters $V_{\max}$, $K_{\mathrm{OP}}$ and $K_{\mathrm{I}}$ (i.e. $\theta_5$, $\theta_6$, $\theta_7$) appear in the same function $V_{\min}$. Their respective impacts on the dynamics are therefore difficult to capture which explain why they are not well identified.

For models of this kind, it is more common to adress the problem of calibration rather than parameter estimation. Calibration consists in finding a set of parameters for which the simulated outputs are close to the observed data. It would require a more detailed analysis, such as that performed in Piccioni et al. [2022]). For illustration, Figure 2 shows one example of the reconstructed time series of phytoplankton and inorganic phosphorus concentrations, obtained using the estimated vector of parameters $\theta$ (estimations obtained with ABC-CNN and ABCD-conformal). Note that accurately reconstructing the dynamics requires simultaneously estimating all parameters correctly. Therefore, the multidimensional coverage and the size of the confidence region are more meaningful than their marginal counterparts.

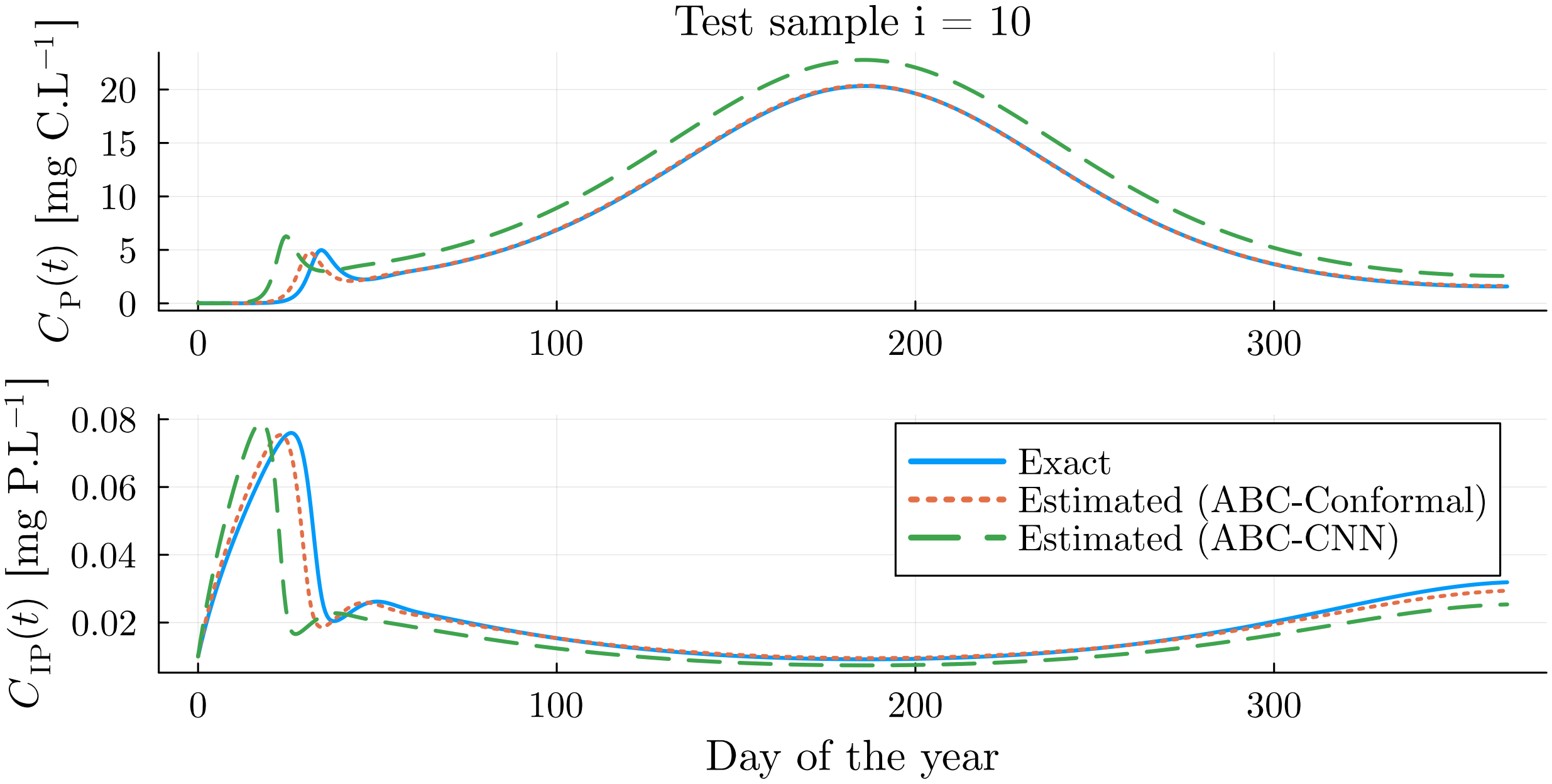

Figure 2: Lake ecosystem dynamics example: reconstructions of time series of phytoplankton and inorganic phosphorus, using the estimated $\theta$ parameters (estimations obtained with ABC-CNN and ABCD-conformal), for the $10^{th}$ test sample.

# 5 Discussion and conclusion

In this article we propose a new ABC method that combines several approaches: the ABC framework, Neural Networks with Monte Carlo Dropout and a conformal procedure. This method is free of any summary statistic, distance, or tolerance threshold. Indeed, it does not require samples to be compared. It behaves well in multidimensional settings, and provides exact non-asymptotic confidences sets.

In practice, this method is computationally efficient, and obtains good results. We test the method on four examples and compare its performances with other approaches: standard ABC, ABC-SMC, ABC-RF and ABC-CNN. ABCD-conformal often outperforms the other methods. Even though ABC-SMC, ABC-CNN or ABC-RF may sometimes achieve better

NMAE results in certain dimensions, we find that the performance of ABCD-conformal is consistently good, among the best if not the best across all parameter dimensions. Regarding ABC-RF, it produced very good results in the unidimensional example, better than ABCD-conformal. However, this algorithm is not adapted for multidimensional parameter estimation. As for ABC-CNN, in low dimensions this method provides good NMAE and is relatively easier to train than ABCD-Conformal. However, it still requires comparing samples using a threshold, which is difficult to calibrate. Moreover, in our experiments, it most often failed to correctly capture the posterior distribution, i.e. its confidence intervals were either too narrow (poor coverage) or, in some cases, too wide, as observed in the lake example (section 4.3). In contrast with the other methods, ABCD-Conformal is the only one with guaranteed coverages for confidence sets. A big advantage of ABCD-Conformal in our practice, is that it always gives both a good estimation accuracy and a good marginal frequentist coverage, which is not always the case for the other methods. In particular in large dimensions, as observed in the lake example (section 4.3), it does not suffer as much as other methods of the curse of dimensionality, providing the best NMAE (within a few percent) and the smallest confidence regions with good coverage. It is an alternative to other methods when there is no obvious summary statistic. The choice of the summary statistics is replaced by the choice of a network architecture. This choice can be guided by common Deep Learning architectures (imaging, time series, ...) and the massive associated literature. We used an almost identical architectures in the four tested examples, i.e. some CNNs with MaxPool to analyse the input features and a few Dense layers to transform back to the parameter space. This is a very different strategy than methods with summary statistics that cannot be applied to different examples. Last but not least, the ABCD-conformal is an amortized procedure, in the sense that once the network has been trained on the reference table and the conformal quantile has been obtained from the calibration set, both this network and this quantile can be reused for any new data sample of interest. There is no need to retrain a network, to recalculate a quantile, or run additional simulations. This nice feature holds for any method that relies on a reference table (such as standard ABC, ABC-CNN or ABC-RF), but not for sequential methods like ABC-SMC.

Nevertheless, the ABCD-conformal approach has certain limitations. One drawback concerns multimodal posteriors. Indeed, the goal of the ABCD-conformal method is to provide point estimates of parameters of interest, along with associated confidence intervals. Unlike standard ABC, ABC-SMC or ABC-CNN, it does not produce an approximation of the full posterior distribution. As a result, in the case of a multimodal posterior, the method is unable to detect such multimodality. Another weakness is that the coverage of the confidence sets is valid marginally, and not conditionally to some values of the parameters of interest. Indeed, on the Lotka-Volterra example (Section 4.2.3) we have seen that the performances of the methods can vary a lot depending on the region of the parameter. However, it should be noted that this limitation applies to all the methods studied. This issue is nonetheless mitigated in the case of ABCD-conformal, thanks to the heuristic uncertainty, which allows the method to adapt its confidence sets according to whether the parameter values are easy or difficult to estimate. In practice, we can note that training neural networks to output both a parameter value and an associated heuristic uncertainty is more challenging than predicting the parameter value alone. Alternative uncertainty quantification methods beyond using the heteroscedastic loss with MC Dropout could be considered (see below). For a more detailed comparison between the five ABC methods used in our examples, the reader can refer to Appendix F. Table 6 gives a summary of this comparison.

The ABCD-Conformal algorithm proposed is promising. However, several improvements and modifications could be considered mostly at the levels of the uncertainty proxy outputted by the neural network and of the conformal procedure. Concerning the neural network, we can imagine using the Dropconnect technique instead of Dropout. These two techniques prevent "co-adaptation" of units in a neural network. Dropout randomly drops hidden nodes, and Dropconnect drops connections (but all nodes can remain partially active). Dropconnect is a generalization of Dropout since there are more possible connections than nodes in a neural network. Another possibility to be explored, could be to use ensembles of neural networks (or deep ensembles) to obtain a random estimation of the parameter of interest and associated uncertainties, see Lakshminarayanan et al. [2017]. As explained by Srivastava et al. [2014c], Dropout can even be interpreted as ensemble model combination. More generally, any method for obtaining an uncertainty associated with the output of a neural network could be used. At the current rate of development of such networks, it seems to us that many methods could be used in the near future. Regarding the conformal procedure, in this article we focused on conformalizing a scalar uncertainty estimate, because the parameter of interest $\theta$ is a vector of scalars. But we can also use conformalized quantile regression (see Angelopoulos and Bates [2023] for a presentation of this procedure). Finally, in the method proposed, we have considered a *split conformal procedure*. This is computationally attractive, as the model needs to be fitted only one time. But it requires having a calibration set, in addition to the training and validation ones (even if in general this set is quite small compared to the training set). *A full conformal procedure* could avoid these extra simulations, at the cost of many more model fits, see Angelopoulos and Bates [2023]. Hence, choosing between split or full conformal procedure could depend on the problem at hand.

# Supplementary materials

**Appendix** It contains additional figures and technical details for the applications, an additional example, pseudocodes and comments for ABC-SMC, ABC-CNN and ABC-RF, and a detailed comparison of the five methods used in the applications. It can be read below the body of this manuscript.

**GitLab repository** The source codes, some datasets and pretrained neural networks to run the examples in section 4 are available in the GitLab repository `https://forge.inrae.fr/mistea/codes_articles/abcdconformal` with the associated `README.md` file. The results are presented as Quarto (or RMardown) notebooks with all instructions to ensure reproducibility.

**Website** Results and html-rendered notebooks can also be directly consulted on the website https://mistea.pages-forge.inrae.fr/codes_articles/abcdconformal/.

**Julia package** The Julia code of all methods used in this article is contained in the Julia package `ABCMethods.jl` accessible in the following Git repository https://github.com/dmetivie/ABCMethods.jl.

Most of the results presented in this article have been obtained using Julia. However, our ABCD-conformal approach was first developed using R. Hence, even if the R results are not presented, R code is provided in the GitLab repository and in the website.

# Acknowledgments

We thank Nicolas Verzelen for introducing us to the conformal approach during a seminar, and for the helpful discussions that followed. We also thank Mathieu Ribatet for the discussions that led to the initiation of this work.

For the applications, we mainly used the Julia programming language [Bezanson et al., 2017] and its scientific computing ecosystem. We acknowledge here the main *open source* packages used for this work. Deep learning components were developed using `Lux.jl` [Pal, 2025], with automatic differentiation via `Zygote.jl` [Innes, 2018] and GPU support through `CUDA.jl` [Besard et al., 2019]. Differential equations and jump processes were modeled using tools from the SciML ecosystem, including `DifferentialEquations.jl` [Rackauckas and Nie, 2017], `JumpProcesses.jl`, and `Catalysis.jl`. Data handling and visualization were supported by `DataFrames.jl` [Bouchet-Valat and Kamiński, 2023], `Distributions.jl` [Besançon et al., 2021], and `Plots.jl` [Christ et al., 2023].

Finally, we are grateful to the co-editor, the associate editor, and two anonymous reviewers for their insightful and constructive comments for improving the quality of the article.

|  | Stand. ABC | ABC-SMC | ABC-CNN | ABCD-Conf. |
|---|---|---|---|---|
| $\theta_1$ parameter | | | | |
| $NMAE(\theta_1)$ | 0.6988 | 0.6298 | 0.3438 | **0.2318** |
| $sd(\lvert\theta_1 - \widehat{\theta_1}\rvert)$ | 0.4150 | 0.4523 | 0.2096 | **0.1782** |
| mean length $CI(\theta_1)$ | 2.5674 | 2.0100 | 1.6073 | **1.1071** |
| median length $CI(\theta_1)$ | 2.6670 | 2.3142 | 1.6073 | **1.0822** |
| coverage $CI(\theta_1)$ | 93.9% | 91.6% | 95.4% | **96.0%** |
| $\theta_2$ parameter | | | | |
| $NMAE(\theta_2)$ | 0.6469 | 0.6526 | 0.3540 | **0.3028** |
| $sd(\lvert\theta_2 - \widehat{\theta_2}\rvert)$ | 0.3992 | 0.3997 | **0.2293** | 0.2411 |
| mean length $CI(\theta_2)$ | 2.3185 | 2.4052 | 1.6189 | **1.2964** |
| median length $CI(\theta_2)$ | 2.4200 | 2.5227 | 1.6079 | **1.1795** |
| coverage $CI(\theta_2)$ | 92.6% | 93.7% | **95.1%** | 94.3% |
| $\theta_3$ parameter | | | | |
| $NMAE(\theta_3)$ | 0.5355 | 0.5295 | **0.2891** | 0.2957 |
| $sd(\lvert\theta_3 - \widehat{\theta_3}\rvert)$ | 0.3594 | 0.4037 | **0.1968** | 0.2401 |
| mean length $CI(\theta_3)$ | 1.9546 | 1.5897 | 1.5722 | **1.2270** |
| median length $CI(\theta_3)$ | 1.9927 | 1.5891 | 1.5667 | **1.1233** |
| coverage $CI(\theta_3)$ | 90.7% | 86.3% | **96.3%** | 94.7% |
| $\theta_4$ parameter | | | | |
| $NMAE(\theta_4)$ | 0.5812 | 0.5674 | 0.2928 | **0.2889** |
| $sd(\lvert\theta_4 - \widehat{\theta_4}\rvert)$ | 0.3770 | 0.3930 | **0.2130** | 0.2314 |
| mean length $CI(\theta_4)$ | 2.1368 | 1.7773 | 1.5487 | **1.2091** |
| median length $CI(\theta_4)$ | 2.2571 | 1.8893 | 1.5386 | **1.0860** |
| coverage $CI(\theta_4)$ | 92.8% | 86.4% | **94.2%** | 93.0% |
| $\theta_5$ parameter | | | | |
| $NMAE(\theta_5)$ | 0.8376 | 0.7146 | 0.6640 | **0.5889** |
| $sd(\lvert\theta_5 - \widehat{\theta_5}\rvert)$ | 0.4677 | 0.4640 | 0.4505 | **0.3907** |
| mean length $CI(\theta_5)$ | 2.8402 | 2.3203 | **1.9357** | 2.3954 |
| median length $CI(\theta_5)$ | 2.8960 | 2.4176 | **1.9476** | 2.4514 |
| coverage $CI(\theta_5)$ | 90.2% | 88.4% | 83.6% | **95.2%** |
| $\theta_6$ parameter | | | | |
| $NMAE(\theta_6)$ | 0.7596 | 0.6933 | 0.7396 | **0.6555** |
| $sd(\lvert\theta_6 - \widehat{\theta_6}\rvert)$ | 0.4392 | 0.4596 | 0.4870 | **0.4180** |
| mean length $CI(\theta_6)$ | 2.5613 | 2.1558 | **1.9344** | 2.5650 |
| median length $CI(\theta_6)$ | 2.6478 | 2.3235 | **1.9508** | 2.9140 |
| coverage $CI(\theta_6)$ | 90.0% | 86.9% | 78.6% | **95.4%** |
| $\theta_7$ parameter | | | | |
| $NMAE(\theta_7)$ | 0.8242 | 0.6419 | 0.4766 | **0.4041** |
| $sd(\lvert\theta_7 - \widehat{\theta_7}\rvert)$ | 0.4709 | 0.3925 | 0.3364 | **0.3133** |
| mean length $CI(\theta_7)$ | 2.7555 | 2.2553 | **1.7518** | 1.7600 |
| median length $CI(\theta_7)$ | 2.8220 | 2.4063 | 1.7186 | **1.6961** |
| coverage $CI(\theta_7)$ | 90.2% | 88.1% | 90.4% | **94.4%** |
| $\theta_8$ parameter | | | | |
| $NMAE(\theta_8)$ | 0.4623 | **0.1661** | 0.2506 | 0.1750 |
| $sd(\lvert\theta_8 - \widehat{\theta_8}\rvert)$ | 0.3139 | 0.1740 | 0.1749 | **0.1333** |
| mean length $CI(\theta_8)$ | 2.1980 | **0.7461** | 1.5547 | 0.8071 |
| median length $CI(\theta_8)$ | 2.2147 | **0.6423** | 1.5445 | 0.7914 |
| coverage $CI(\theta_8)$ | 94.6% | 95.5% | **96.6%** | 95.2% |
| $\theta_9$ parameter | | | | |
| $NMAE(\theta_9)$ | 0.3888 | 0.5478 | 0.2545 | **0.1413** |
| $sd(\lvert\theta_9 - \widehat{\theta_9}\rvert)$ | 0.2760 | 0.3773 | 0.1717 | **0.1143** |
| mean length $CI(\theta_9)$ | 1.6639 | 1.8246 | 1.5408 | **0.6059** |
| median length $CI(\theta_9)$ | 1.6985 | 1.8451 | 1.5252 | **0.5815** |
| coverage $CI(\theta_9)$ | 94.9% | 89.9% | **96.9%** | 95.0% |
| $\theta$ parameter, 9-dimensional | | | | |
| mean length $CE(\theta)$ | 2235 | **57.12** | 823.3 | 110.5 |
| median length $CE(\theta)$ | 11683 | **14.70** | 765.0 | 102.8 |
| coverage $CE(\theta)$ | 90.3% | 73.4% | 87.9% | **95.4%** |

Table 5: Lake system dynamics example: for each method and each component of $\theta$, we give the NMAE, $sd(|\theta - \widehat{\theta}|)$, the mean and median lengths of the confidence or credible intervals and the coverage. For each method the mean and median volumes of multidimensional confidence or credible ellipses are also given, as well as the associated coverage. The best values for each line are highlighted in bold.

| **Point of comparison** | **Standard ABC** | **ABC-SMC** | **ABC-CNN** | **ABC-RF** | **ABCD-Conf** |
|---|---|---|---|---|---|
| Approx. of the whole posterior or of transforms of interest | whole posterior | whole posterior | whole posterior | transforms of interest | transforms of interest |
| No need of relevant summary stat | ✗ | ✗ | ✓ | ✗ | ✓ |
| No need of a distance | ✗ | ✗ | ✓ | ✓ | ✓ |
| No need of a tolerance threshold | ✗ | ✗ | ✗ | ✓ | ✓ |
| Deal with multidimensional parameters | ✓ | ✓ | ✓ | ✗ | ✓ |
| No need to choose network or random forest architecture | ✓ | ✓ | ✗ | ✗ | ✗ |
| Datasets needed | 1: training | Simulations made sequentially | 2: training, validation | 1: training | 3: training, validation, calibration |
| Justification of the method, guarantees | asymptotic under conditions | asymptotic under conditions | asymptotic conditions too difficult to check | asymptotic conditions too difficult to check | non asymptotic no condition |
| Adapted for different types of data and high-dim data | difficult ✗ | difficult ✗ | ✓ | difficult ✗ | ✓ |
| Computing time | difficult to compare, it depends on examples. | | | | |

Table 6: Summary table comparing the five methods: standard ABC, ABC-SMC, ABC-CNN, ABC-RF and ABCD-Conformal. A detailed comparison of these methods can be found in appendix F.

The authors report there are no competing interests to declare.

# A Application: Moving Average 2

Complementing Section 4.1 of the article, additional figures are given for illustration purposes. Note that all these figures can be find in Notebooks associated with this publication.

- Figure A.1 shows, for a test sample, the true parameter $\theta$, as well as estimated $\hat{\theta}$ using standard ABC, ABC-CNN and ABCD-conformal and associated confidence ellipses.

- Figure A.2 illustrates that the true value of $\theta$ is close to the associated exact posterior mean. This last one is calculated from explicit formulas from Marin and Robert [2007][Chapter 5] and requires the inversion of a large matrix, which is numerically costly and a source of possible additional numerical error.

# B Application: Lotka Volterra

Complementing Section 4.2 of the article, additional figures are given for illustration purposes. Note that all these figures can be find in Notebooks associated with this publication.

- Figure B.1 shows for the test sample 997, estimated parameter $\theta$ with associated 3D confidence ellipsoids obtained using standard ABC, ABC-CNN and ABCD-conformal, and true value of $\theta$.

- Figure B.2 shows for the test sample 667, estimated parameter $\theta$ with associated 3D confidence ellipsoids obtained using standard ABC, ABC-CNN and ABCD-conformal, and true value of $\theta$.

# C Application: Lake Ecosystem dynamics

## C.1 Detailed description of the model

The system we are considering here corresponds to a lake ecosystem for which the time evolution of the phytoplankton and the inorganic phosphorus concentrations is represented through the following equations:[1]

$$\frac{\mathrm{d}C_{\mathrm{P}}}{\mathrm{d}t} = \mu(C_{\mathrm{IP}}, T)C_{\mathrm{P}} - d_{\mathrm{P}}C_{\mathrm{P}} \tag{16}$$

$$\frac{\mathrm{d}C_{\mathrm{IP}}}{\mathrm{d}t} = V_{\min}(C_{\mathrm{OP}}, C_{\mathrm{IP}}, T) - a_{\mathrm{pc}}\mu(C_{\mathrm{IP}}, T)C_{\mathrm{P}} + (1 - f_{\mathrm{OP}})a_{\mathrm{pc}}d_{\mathrm{P}}C_{\mathrm{P}} \tag{17}$$

where $C_{\mathrm{P}}$ and $C_{\mathrm{IP}}$ respectively represent the phytoplankton concentration [mg C.L$^{-1}$] and the inorganic phosphorus concentration [mg P.L$^{-1}$], $C_{\mathrm{OP}}$ is the organic phosphorus concentration

[1] mg, P, C and L stand for milligram, phosphorus, carbon and liter respectively.

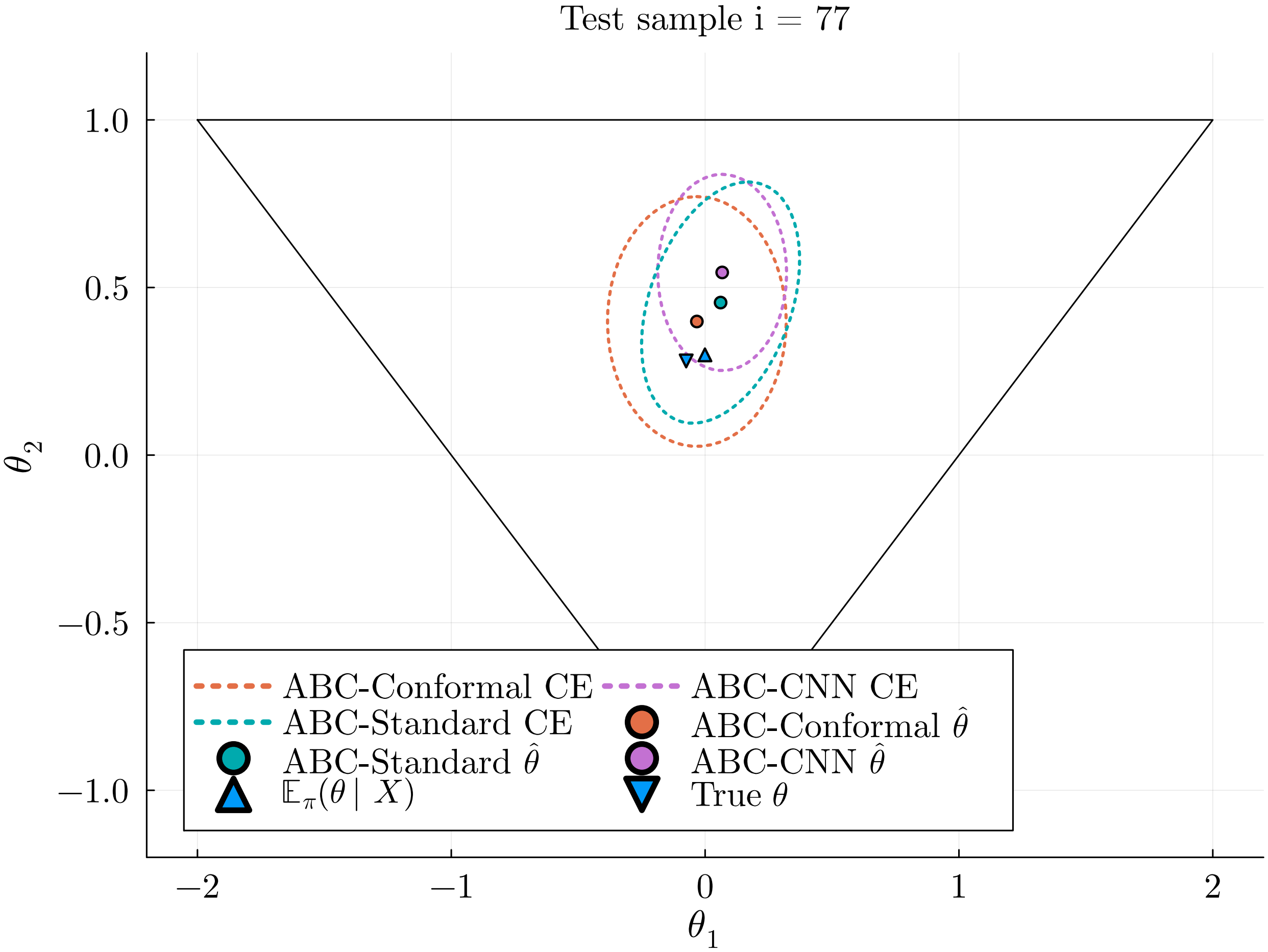

Figure A.1: Moving Average 2 example: for the 77th test sample, the true value of $\theta$ is represented (up-triangle), as well as the estimated $\hat{\theta}$ using standard ABC, ABC-CNN and ABCD-conformal and associated confidence ellipses (circles and ellipses). The exact posterior mean is also represented (down triangle), for comparison with the true value.

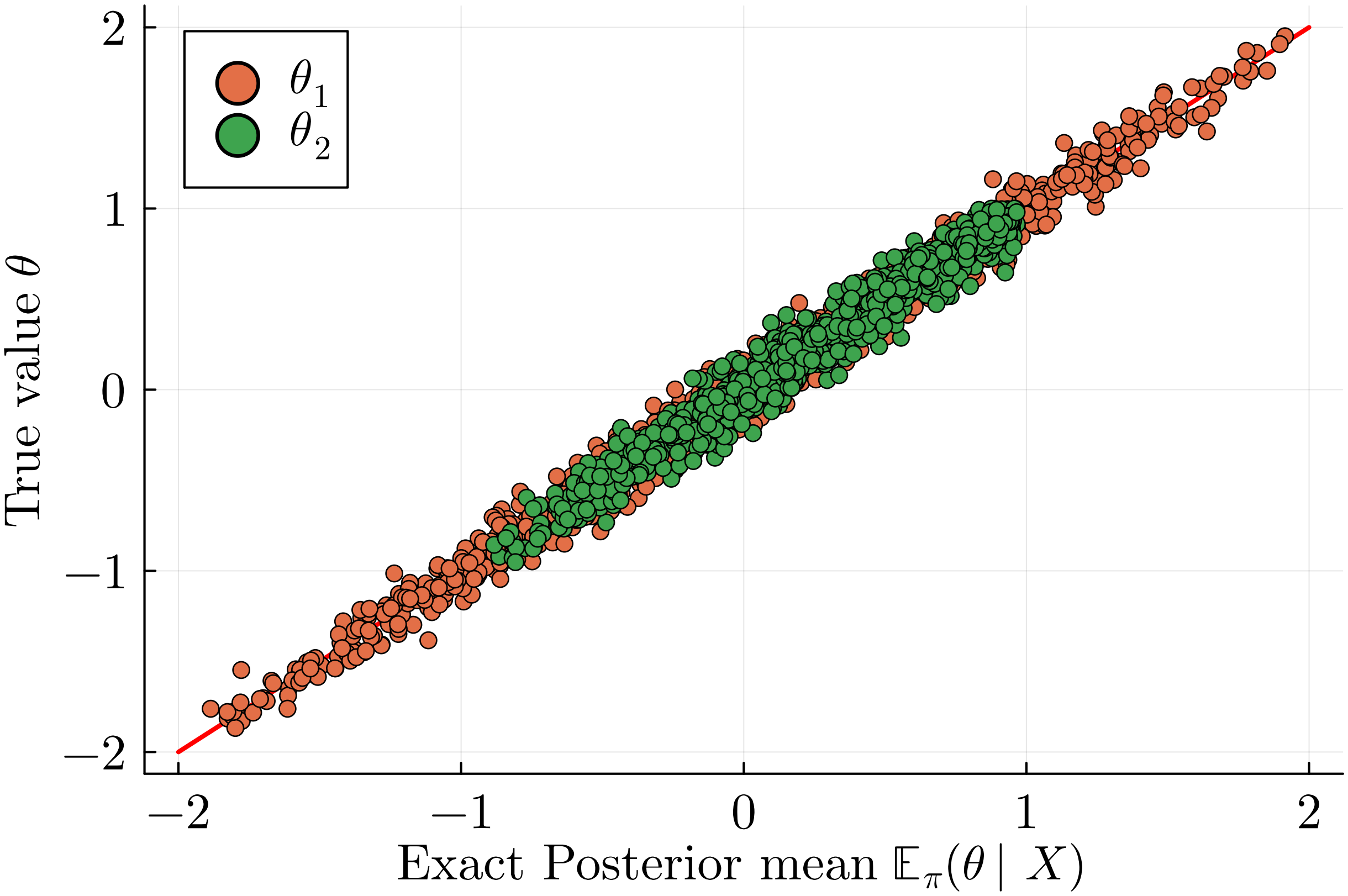

Figure A.2: Moving Average 2 example: each red (resp. green) dot represents the exact posterior mean $\mathbb{E}(\theta_1|\mathbf{x})$ (resp $\mathbb{E}(\theta_2|\mathbf{x})$) in the x-axis, against the true value of $\theta_1$ (resp. $\theta_2$) in the y-axis, for each of the 1000 test samples.

[mg P.L$^{-1}$] that is assumed to evolve slowly compared to the other variables, and that is given, due to the law of conservation of mass, by:

$$C_{\mathrm{OP}} = \varepsilon(K_0 - a_{\mathrm{pc}}C_{\mathrm{P}} - C_{\mathrm{IP}}) \tag{18}$$

where $K_0 = a_{\mathrm{pc}}C_{\mathrm{P},0} + \frac{1}{\varepsilon}C_{\mathrm{OP},0} + C_{\mathrm{IP},0}$ [mg P.L$^{-1}$] represents the initial total phosphorus concentration in the water (which is conserved) and $C_{\mathrm{P},0}$ [mg C.L$^{-1}$], $C_{\mathrm{IP},0}$ [mg P.L$^{-1}$] and $C_{\mathrm{OP},0}$ [mg P.L$^{-1}$] are the initial concentrations of phytoplankton, inorganic phosphorus and organic phosphorus in the water.

In the model, the parameter $d_{\mathrm{P}}$ represents the death rate of phytoplankton [day$^{-1}$], $f_{\mathrm{OP}}$ is the fraction of dead phytoplankton going to the organic phosphorus pool [-], $a_{\mathrm{pc}}$ is the phosphorus-to-carbon ratio of the phytoplankton [mg P.(mg C)$^{-1}$] and $\varepsilon$ [-] is an adimensional parameter reflecting the slow evolution of organic phosphorus compared to the other variables.

The time-dependent temperature of the water $T$ is assumed to be a sinusoidal function

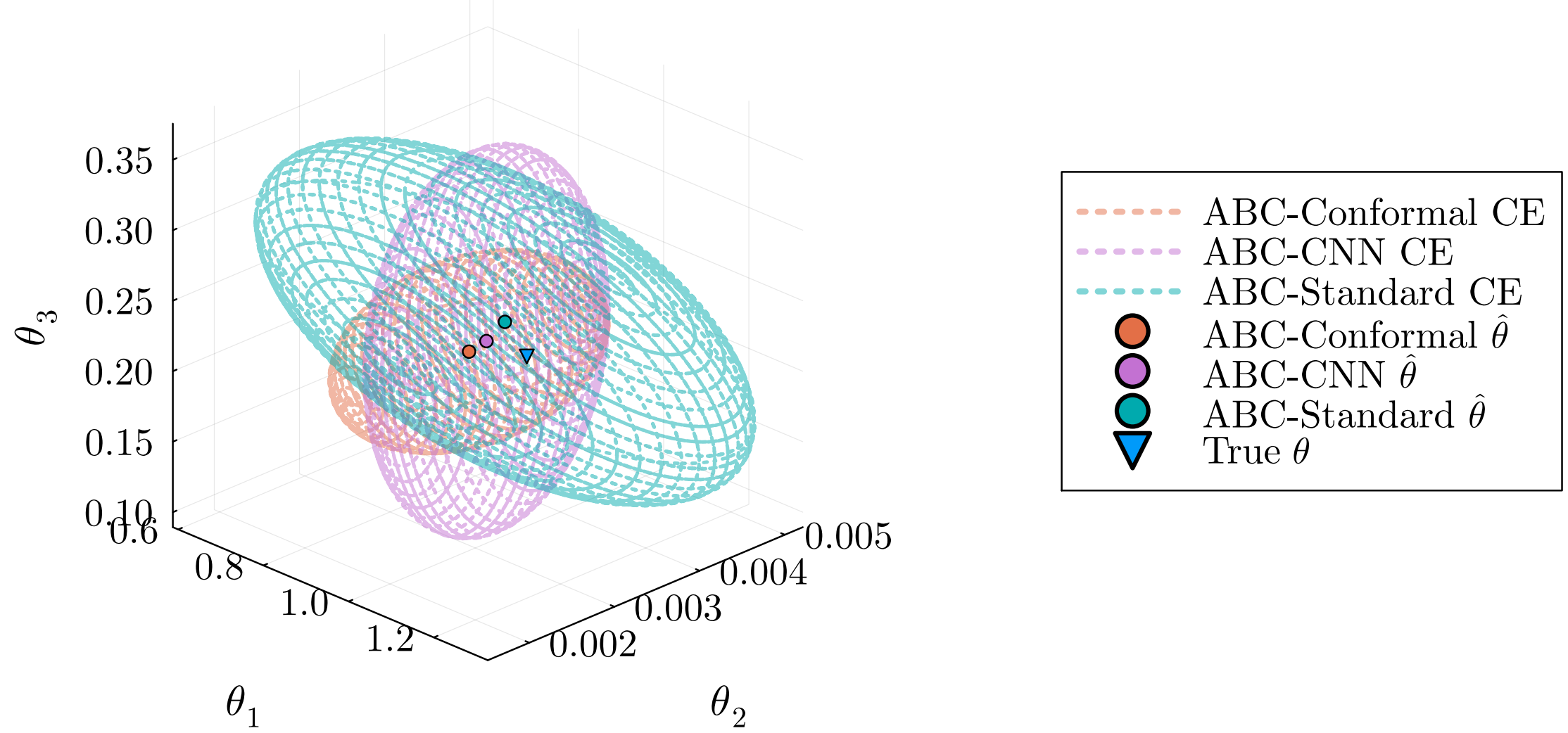

Figure B.1: Lotka-Volterra example: for the test sample 997, estimated parameter $\theta$ with associated 3D confidence ellipsoids obtained using standard ABC, ABC-CNN and ABCD-conformal, and true value of $\theta$.

of the form:

$$T(t) = \left(\cos\left(\frac{\pi}{f}t + \pi\right) + 1\right)\frac{T_{\max} - T_{\min}}{2} + T_{\min} \tag{19}$$

where $T_{\max}$ [$^\circ$ C] and $T_{\min}$ [$^\circ$ C] are the maximal and minimal temperatures reached by the water temperature during the year and $f$ [day] is the period of the sinusoidal function.

The growth rate of the phytoplankton $\mu$ is expressed by:

$$\mu(C_{\mathrm{IP}}, T) = \mu_{\max}\frac{C_{\mathrm{IP}}}{K_{\mathrm{IP}} + C_{\mathrm{IP}}}\theta_{gr}^{T-20} \tag{20}$$

where $\mu_{\max}$ is the maximal growth rate of the inorganic phosphorus uptake [day$^{-1}$], $K_{\mathrm{IP}}$ is the half saturation constant of the phytoplankton growth function [mg P.L$^{-1}$], and $\theta_{gr}$ is the temperature coefficient for phytoplankton growth [-].

Finally the mineralization speed of the organic phosphorus into inorganic phosphorus $V_{\min}$ is given by:

$$V_{\min}(C_{\mathrm{OP}}, C_{\mathrm{IP}}, T) = V_{\max}\frac{C_{\mathrm{OP}}}{K_{\mathrm{OP}}\left(1 + \frac{C_{\mathrm{IP}}}{K_{\mathrm{I}}}\right) + C_{\mathrm{OP}}}\theta_{\min}^{T-20} \tag{21}$$

with $V_{\max}$ the maximal mineralization rate [mg P.L$^{-1}$.day$^{-1}$], $K_{\mathrm{OP}}$ the half saturation constant of the mineralization function [mg P.L$^{-1}$], $K_{\mathrm{I}}$ the inhibition coefficient of the enzymatic reaction [mg P.L$^{-1}$] and $\theta_{\min}$ the temperature coefficient for mineralization [-].

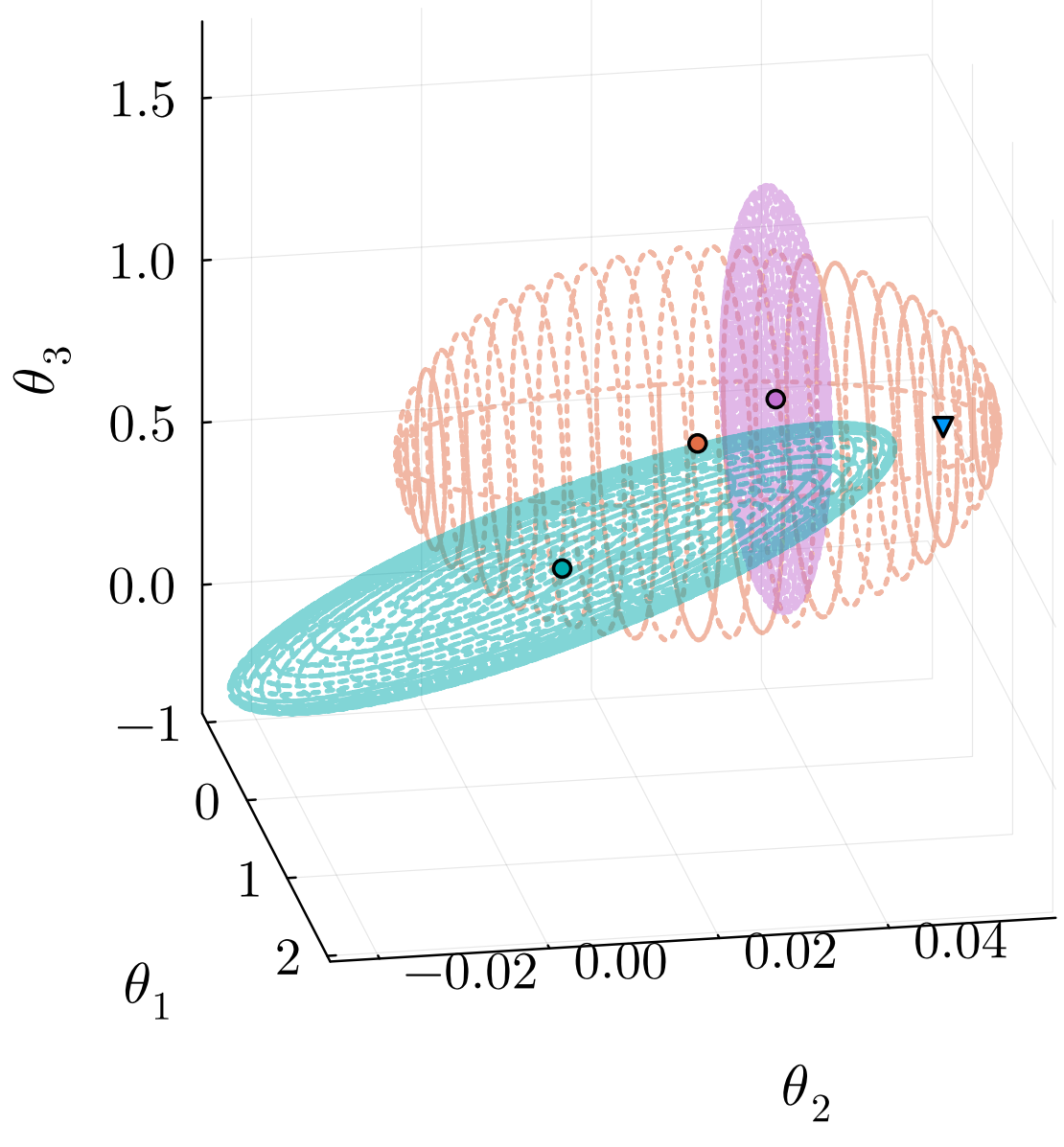

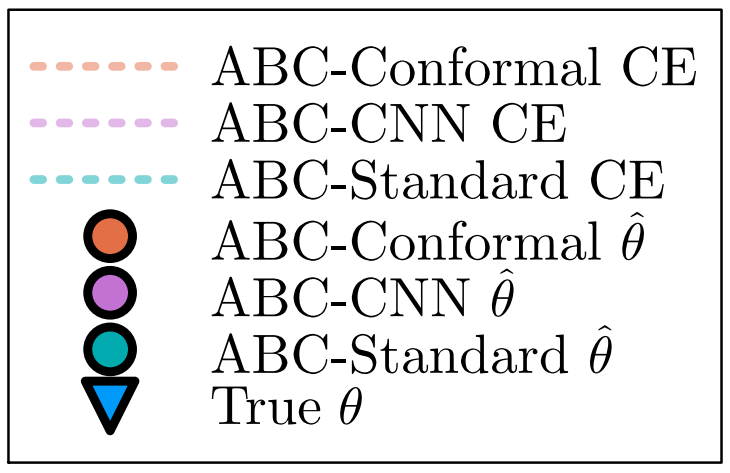

Figure B.2: Lotka-Volterra example: for the test sample 667, estimated parameter $\theta$ with associated 3D confidence ellipsoids obtained using standard ABC, ABC-CNN and ABCD-conformal, and true value of $\theta$.

In the model, there are 15 parameters and 2 initial conditions. Among the 15 parameters, the following 6 parameters have a fixed value and are not included in the estimation process:

$$a_{\text{pc}} = 0.00625\,[\text{mg P.(mg C)}^{-1}],\ T_{\text{min}} = 5\,[^\circ C],\, T_{\text{max}} = 30\,[^\circ C],$$
$$f = \frac{365}{2}\,[\text{day}],\, \varepsilon = 10^{-6}\,[\text{-}],\, C_{\text{OP},0} = 1\,[\text{mg P.L}^{-1}].$$

The objective is to estimate the 9 remaining parameters that is the vector $\theta$ given by: $\theta = (\mu_{\text{max}}, K_{\text{IP}}, d_{\text{P}}, f_{\text{OP}}, V_{\text{max}}, K_{\text{OP}}, K_{\text{I}}, \theta_{\text{min}}, \theta_{\text{gr}})$

The initial conditions are fixed to $C_{\text{P},0} = 0.01$ [mg P.L$^{-1}$], and $C_{\text{IP},0} = 0.01$ [mg P.L$^{-1}$]. A dataset corresponds to observations of state $(C_{\text{P}}, C_{\text{IP}})$ at times $0, 0.5, 1, \ldots, 364.5$ days (i.e. on a one year time-scale, every 12 hours). The prior distributions are independent uniforms $\mathcal{U}[\theta_i^m, \theta_i^M]$ for each parameters $\theta_i$ of $\theta$, with $\theta^m = (1, 0.0009, 0.03, 0, 0.05, 0.03, 0, 1, 1)$ and $\theta^M = (3, 0.052, 0.7, 1, 0.22, 3, 0.04, 1.1, 1.1)$.

## C.2 Additional figures

Complementing Section 4.3 of the article, additional figures are given for illustration purposes. Note that all these figures can be find in Notebooks associated with this publication.

Figures C.1, C.2, C.3 and C.4 show for each component of $\theta$, the estimated values against the true value for the 1000 test samples, as well as associated confidence sets. The estimations are obtained using standard ABC, ABC-SMC, ABC-CNN and ABCD-conformal respectively.

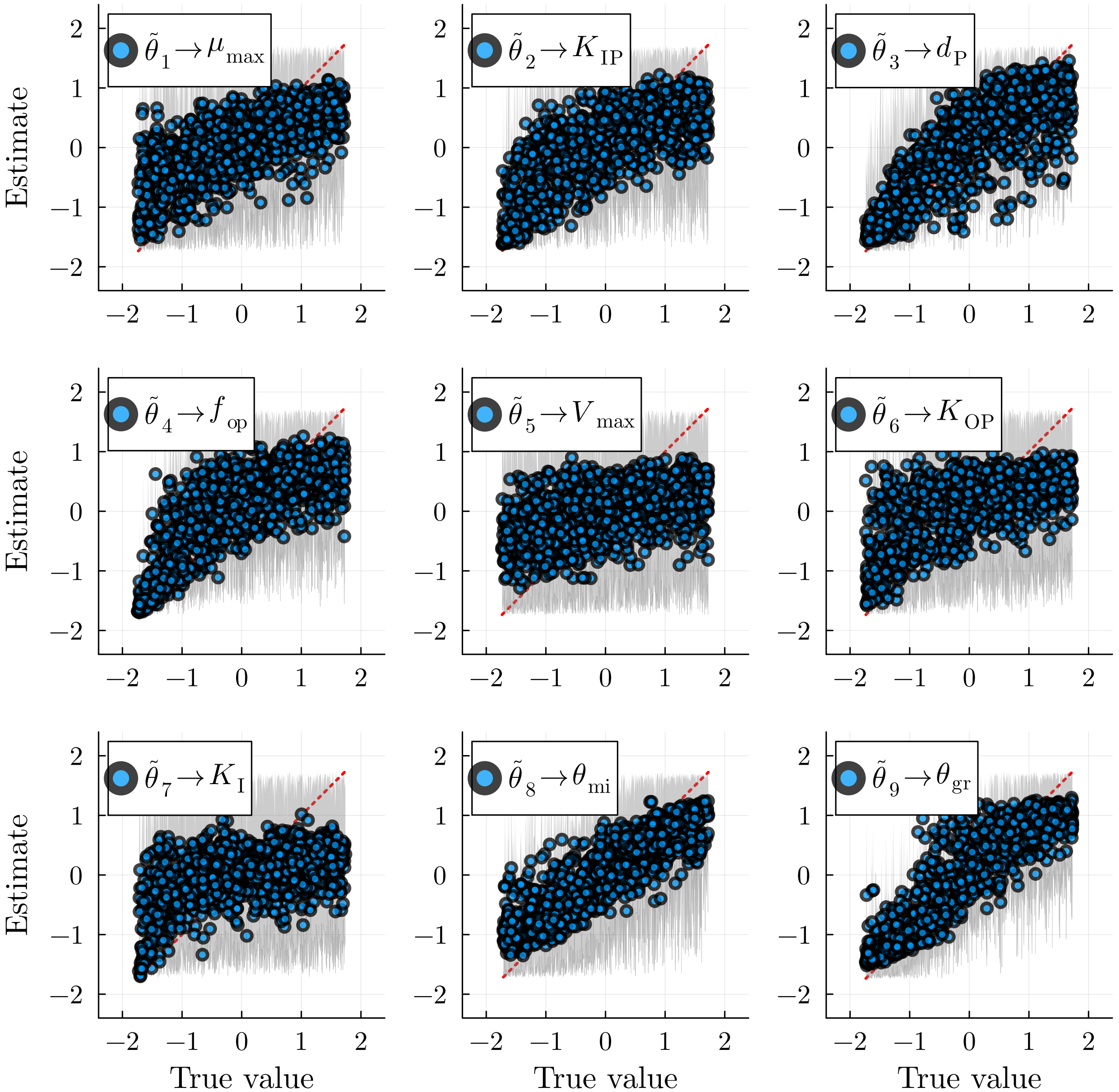

Figure C.1: Lake ecosystem dynamics example: for each component of normalized $\theta$ denoted $\tilde{\theta}$, estimated values against true values, for 1000 test samples, as well as associated confidence intervals (in gray, for the overall uncertainty). The estimated values are obtained using standard ABC.

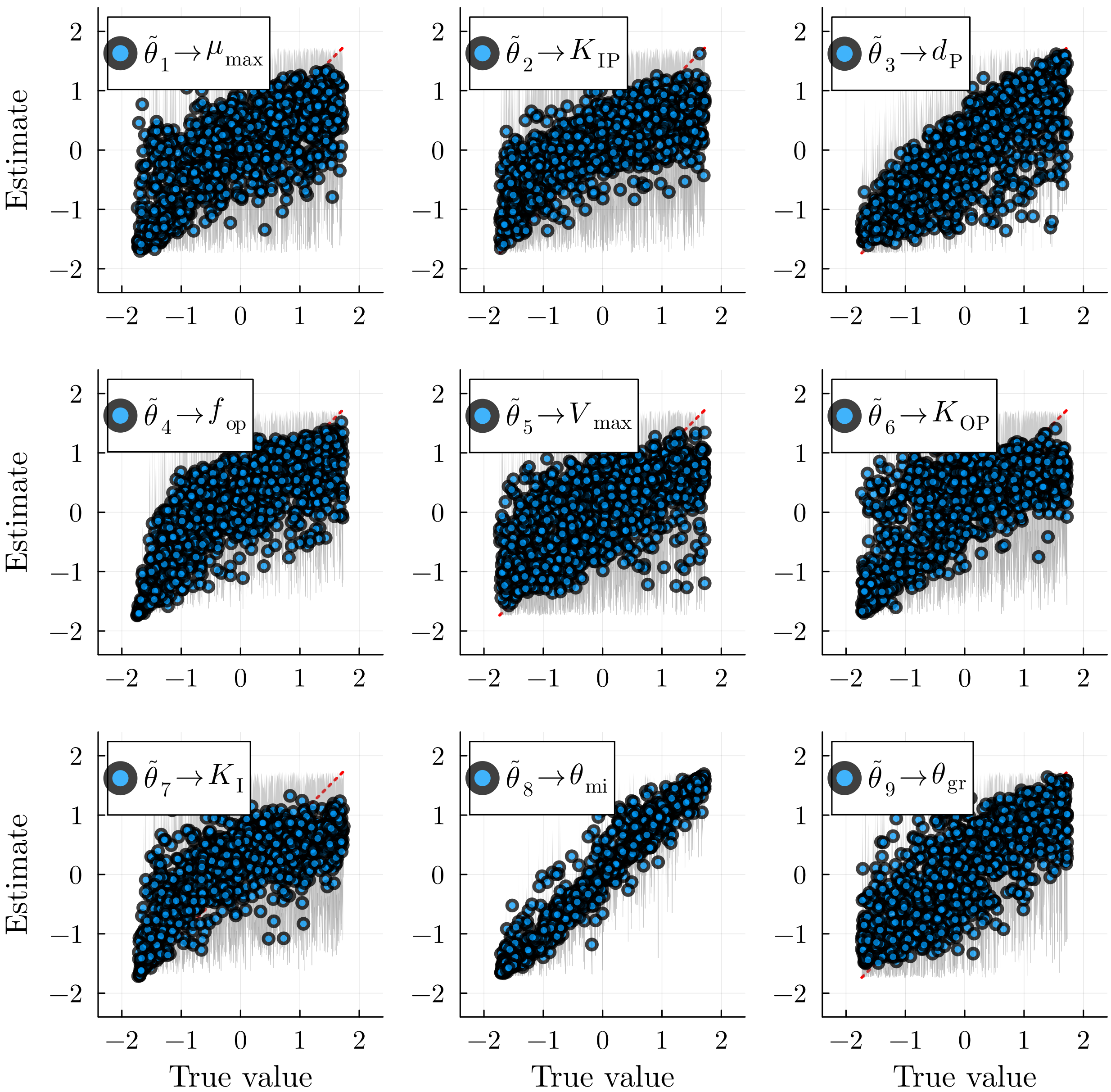

Figure C.2: Lake ecosystem dynamics example: for each component of normalized $\theta$ denoted $\tilde{\theta}$, estimated values against true values, for 1000 test samples, as well as associated confidence intervals (in gray, for the overall uncertainty). The estimated values are obtained using ABC-SMC.

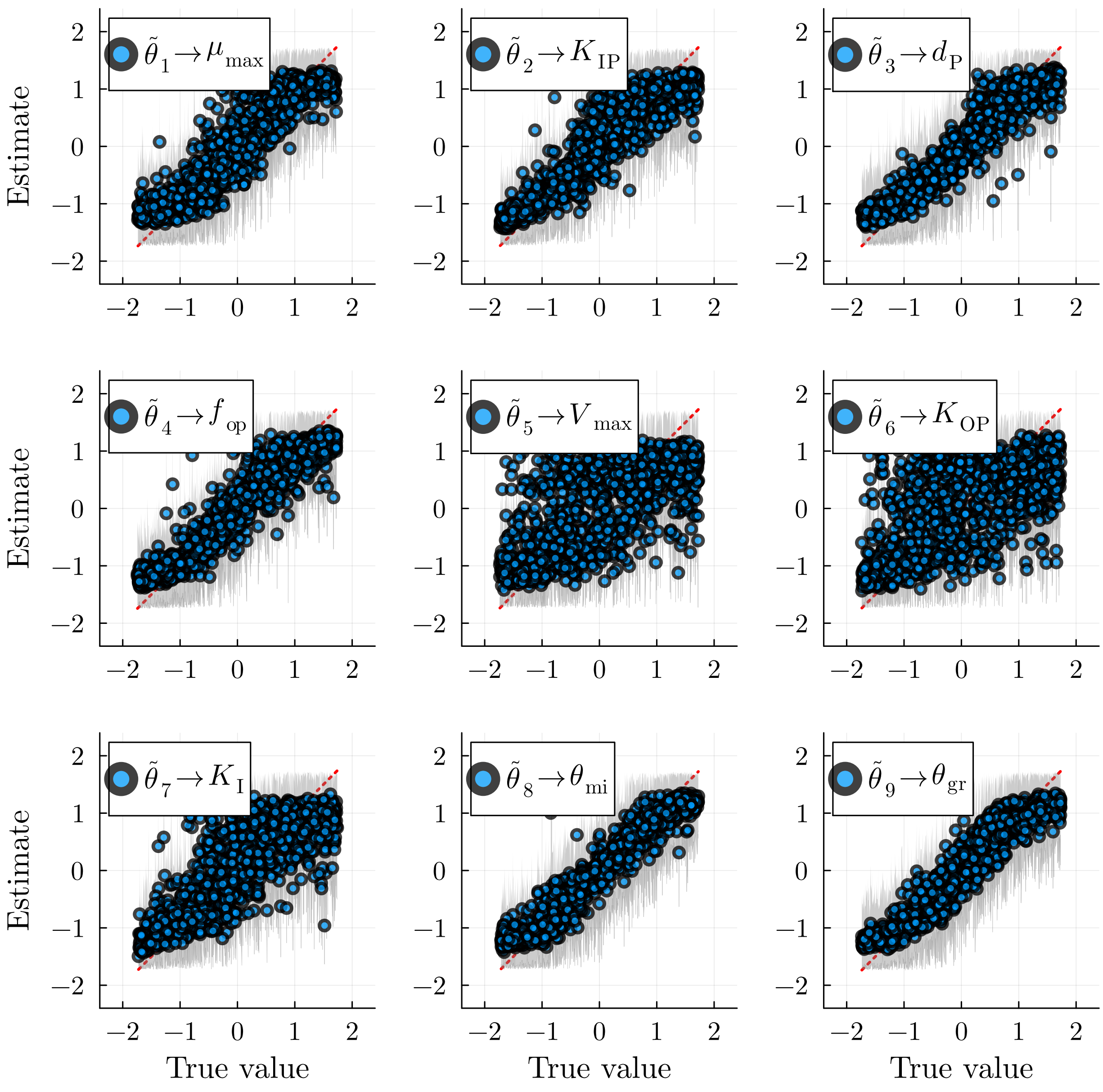

Figure C.3: Lake ecosystem dynamics example: for each component of normalized $\theta$ denoted $\tilde{\theta}$, estimated values against true values, for 1000 test samples, as well as associated confidence intervals (in gray, for the overall uncertainty). The estimated values are obtained using ABC-CNN.

# D Application: Gaussian random field

## D.1 The model

We study stationary isotropic Gaussian random field on the domain $[0, 5] \times [0, 5]$ with a regular grid of size $100 \times 100$, with exponential covariance functions [Wood and Chan, 1994].

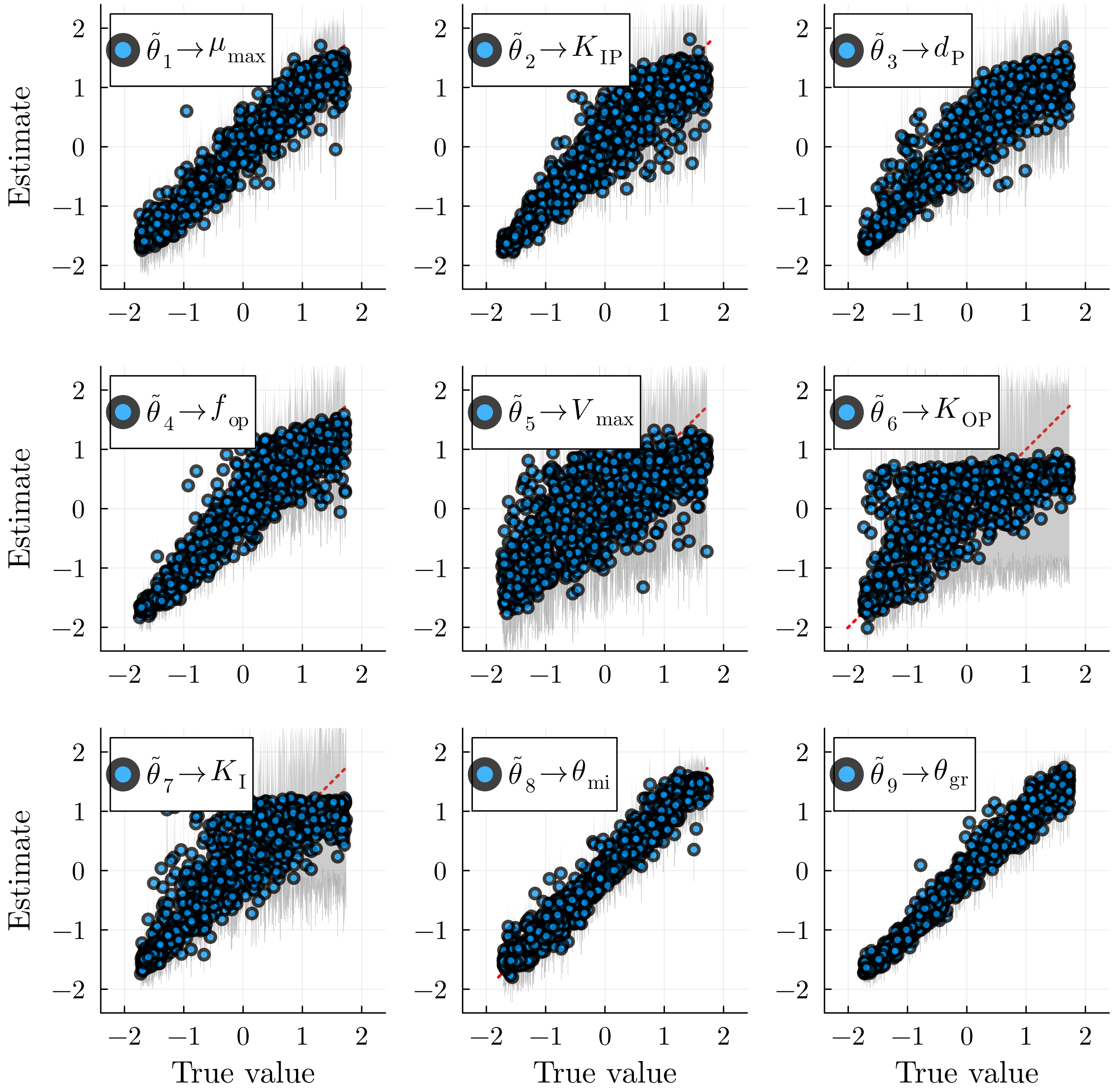

Figure C.4: Lake ecosystem dynamics example: for each component of normalized $\theta$ denoted $\tilde{\theta}$, estimated values against true values, for 1000 test samples, as well as associated confidence intervals (in gray, for the overall uncertainty). The estimated values are obtained using ABCD-conformal.

This covariance between two points $z_1$ and $z_2$ is given by:

$$C(z_1, z_2) = \exp\left( \left|\left| \frac{z_1 - z_2}{\theta} \right|\right|^2 \right),$$

with $\theta$ the range (or scale) parameter, which is the unknown parameter. We assume that the prior distribution on $\theta$ is the uniform distribution between 0 and 1.

## D.2 Algorithm parametrization

Number of used samples are given in Table 1 of the article. For the standard ABC and the ABC-RF, we use two summary statistics: the Moran's $I$ statistics from lag 1 to 5 (that is the Moran's correlogram from lags 1 to 5), and the semi-variogram up to a distance of 20 (15 values kept per variogram) (see Cliff and Ord [1981] for these notions). We assume the spatial weights matrix to be row-standardized, and the neighbors of a pixel being the 4-nearest pixels. As all Gaussian fields are simulated on the same grid, the variograms of the different fields are calculated at exactly the same distances, so they are comparable. For the standard ABC, the distance used to compare two Gaussian random fields is the sum of the quadratic distance between their Moran's correlograms and of the quadratic distance between their semi-variograms.

Concerning the architectures of neural networks for ABC-CNN and ABCD-conformal, we used 3 convolutional 2D layers having 32, 64 and 64 neurons, kernels of size 3 and the `relu` activation function, followed by max-pooling for the first two layers and by a flatten layer for the third one. Then, 2 dense layers having both 64 neurons and the `relu` activation function. The last layer to obtain the output of dimension 1 is linear. The raw samples given as inputs of the neural networks are the 2D Gaussian random fields, and the associated outputs are the estimations of the parameter $\theta$.

Concerning ABCD-conformal, we look at its performances using two different measures of uncertainty: the overall uncertainty and the epistemic uncertainty, both returned by the Dropout procedure.

**Remark 4.** *This example is the only one where we use the ABC-RF which can be applied only on uni-dimensional parameters. For the other examples with multidimensional parameters, we only considered methods that can deal with multidimensionality.*

**Remark 5.** *In this example, we do not show results of ABC-SMC. The ABC-SMC was implemented, but using R the algorithm took too long to be run for a sufficient number of iterations and satisfactory population sizes. Indeed, ABC-SMC is the only method that do not rely on a reference table and by consequence requires many simulations for each test sample, see section E.1 for a discussion about that.*

## D.3 Results

All indicators detailed in the beginning of Section 4 of the article are presented in Table 7. Figure D.1 shows the predicted parameters $\widehat{\theta}$ estimated by the ABCD-Conformal method against the true values for the 100 test samples, and using gray intervals.

In Table 7, we see that in this example, ABC-RF outperforms the others methods, with the smallest NMAE and $sd(|\theta - \widehat{\theta}|)$, while having a very good coverage, 98%, which is better than the expected coverage of 95%. The standard ABC and ABC-CNN do not give

| | Standard ABC | ABC-CNN | ABC-RF | ABCD-Conformal overall | ABCD-Conformal epistemic |
|---|---|---|---|---|---|
| $NMAE(\theta)$ | 0.0436 | 0.0223 | **0.0151** | 0.0300 | 0.0300 |
| $sd(\lvert\theta - \widehat{\theta}\rvert)$ | 0.0229 | 0.0106 | **0.0076** | 0.0141 | 0.0141 |
| mean length $CI(\theta)$ | 0.1313 | **0.0307** | 0.0405 | 0.0684 | 0.0711 |
| median length $CI(\theta)$ | 0.1341 | **0.0298** | 0.0391 | 0.0637 | 0.0672 |
| coverage $CI(\theta)$ | **100%** | 88% | 98% | 93% | 94% |

Table 7: Two dimensional Gaussian random field: for each method, we give the NMAE, $sd(|\theta - \widehat{\theta}|)$, the mean and median lengths of the confidence or credible intervals and the coverage. These indicators are computed on a test set of size $10^2$. The best values for each line are highlighted in bold.

satisfactory results. Indeed, while having a coverage of 100%, the standard ABC has the largest NMAE and $sd(|\theta - \widehat{\theta}|)$, and the good coverage is due to too large confidence intervals. In the case of the ABC-CNN, like in the previous example, the NMAE, $sd(|\theta - \widehat{\theta}|)$ and the lengths of the confidence sets are quite small, but it is counterbalanced by a too small coverage (88%). Finally, the results given by the ABCD-Conformal approach are satisfactory, with coverages close to 95%. The overall and the epistemic uncertainty give quite similar results in the conformal procedure.

To sum up, the ABC-RF is efficient here on all criteria. ABCD-Conformal and ABC-CNN give good predictions, but ABCD-Conformal is much better on coverages. Standard ABC can be discarded compared to other methods in view of its poor results, except for the coverage.

# E Pseudocodes for ABC-SMC, ABC-CNN and ABC-RF and some comments

## E.1 ABC-SMC

Like the standard ABC algorithm, the ABC-SMC requires summary statistics and a distance to compare two data samples. While standard ABC only requires a single threshold, ABC-SMC needs a sequence of decreasing thresholds (or a way to compute it) at each iteration. Other hyperparameters include the size of the sequential populations $N$, the number of iterations $T$, as well as perturbation samplers (transition kernels) $\mathcal{K}$ used to move from one population to another. Therefore, compared to standard ABC, ABC-SMC involves many more calibration parameters. Generally speaking, ABC-SMC yields better results than standard ABC, but requires a bigger expertise.

That said, assuming an informed user knows how to set up an ABC-SMC algorithm, one major difference between ABC-SMC and all the other methods used in this paper is that simulations must be carried out sequentially. In contrast, standard ABC, ABC-CNN, ABC-

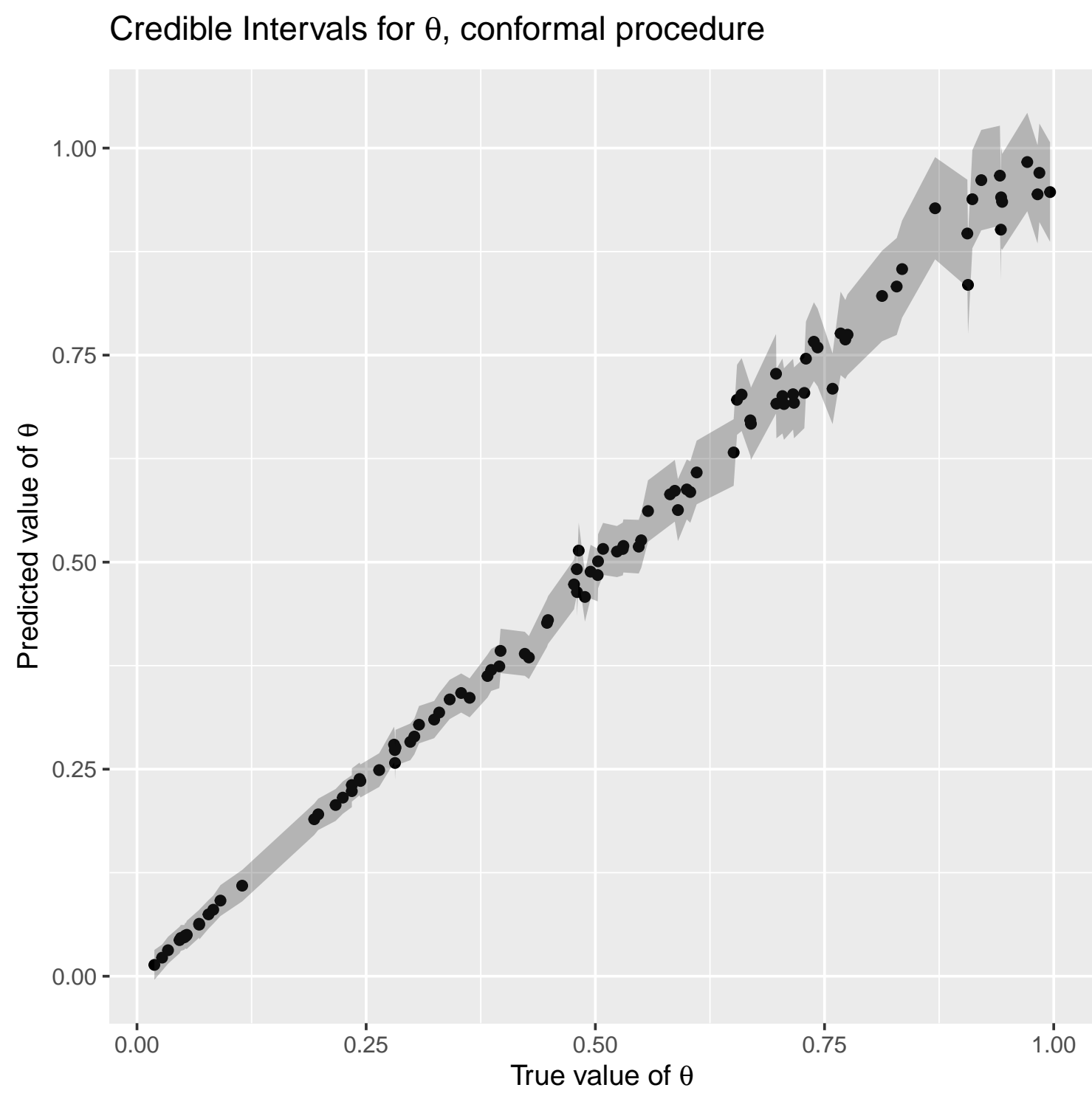

Figure D.1: Two-dimensional Gaussian random field example: predicted parameters $\widehat{\theta}$ estimated by the ABCD-Conformal against the true values for the 100 test samples, and (in gray intervals, for the overall uncertainty).

RF, and ABCD-conformal benefit from the use of a reference table, which allows simulations to be generated in advance, potentially through parallelization. Additionally, the same reference table can be reused to estimate parameters for different data samples of interest. This enables these algorithms to be calibrated and tested much more efficiently. For example, for a test set of 1000 samples, it is not necessary to generate a new reference table for each test sample while the ABC-SMC, on the other hand, requires a new set of simulations for each test sample.

The version of the ABC-SMC algorithm used for the numerical simulations is given below in Algorithm 3. Its implementation is very similar to that of the R package `smfsb` Wilkinson [2024]. There are multiple ways to define a stopping criterion for the algorithm, for instance, one can choose to fix the total number of simulations performed during the process, to prevent excessively long runs (this was our choice). However, in this case, the final tolerance level is not guaranteed. Conversely, the user may decide to fix a final tolerance level that needs to be reached, but in this case, the total number of simulations is not known in advance. This is just one example; many different versions of this algorithm have been proposed since Sisson et al. [2007]; Beaumont et al. [2009] and Del Moral et al. [2012].

To complement the Algorithm 3, here are a few remarks:

- Regarding the transition kernel $\mathcal{K}$ used to perturb a random draw from a previous population (line 10 of Algorithm 3), we chose the classic Gaussian kernel (e.g., Filippi et al. [2013]). At iteration $t$, this kernel is centered on a random draw from the weighted population $(t-1)$ and has a covariance matrix equal to the empirical weighted covariance matrix of this population, $\Sigma_{t-1}$. The importance weights are then given by (with the usual abuse of notation): $\tilde{w}_i^{(t)} = \pi(\theta_i^{(t)}) / \sum_{j=1}^{N} w_j^{(t-1)} \mathcal{N}(\theta_i^{(t)}; \theta_j^{(t-1)}, 2\Sigma_{t-1})$, where $\mathcal{N}(\theta; m, \Sigma)$ denotes the probability density function of a Gaussian distribution with mean $m$ and covariance matrix $\Sigma$ evaluated at $\theta$. Note that we experimented with the popular choice of using a factor of 2 for the covariance matrix [Filippi et al., 2013], with no clear advantage.

- At line 17 of Algorithm 3, the new set $\theta^{(t)}$ is, by construction, of size $N$. In the case of ties for the $N$-th smallest distance, we arbitrarily select from the associated particles $\theta_j$ to ensure the set contains exactly $N$ particles. For each iteration $t$, this way of selecting particles corresponds to a tolerance level $\epsilon_t$ defined by $\Delta_{(N)}$.

- For fair comparisons with other methods, we ensure that the total number of calls to the model $f$ is equal to $N_{\text{train}}$, i.e., $N \times \Psi \times T = N_{\text{train}}$. Note that this means we must generate $N_{\text{train}}$ samples from the model for every test sample, which is significantly more time-consuming than other methods that use a common reference table.

- To guarantee fairness in the Lotka-Volterra example (Section 4.2 of the article), we require the draw at line 12 of Algorithm 3 to yield a valid, non-zero trajectory. If an extinction occurs, the draw is discarded and we return to line 9 of the algorithm.

**Algorithm 3:** Pseudocode for the ABC-SMC sampler [Sisson et al., 2007; Beaumont et al., 2009; Del Moral et al., 2012]

**Input :** A Bayesian parametric model $\{f(\cdot \mid \theta), \pi\}$, a data sample $\mathbf{x}$, the number of samples kept per iteration $N$, the number of steps $T$, summary statistics $\eta : \mathbb{R}^d \to \mathbb{R}^l$, a (pseudo)-distance $d$ on $\mathbb{R}^l$, a perturbation sampler $\mathcal{K}_t$ and an integer factor $\Psi > 0$.

**Output:** A weighted set $(\theta_1^{(T)}, \tilde{w}_1^{(T)}), \ldots, (\theta_N^{(T)}, \tilde{w}_N^{(T)})$ whose distribution is approximately $\pi(\theta \mid \mathbf{x})$

1 **Initial population** :
2 **for** $i \leftarrow 1$ **to** $N$ **do**
3     Set $\theta_i^{(0)} \sim \pi$;
4     Set $w_i^{(0)} := 1/N$.
5 **end**
6 **Sequential populations** :
7 **for** $t \leftarrow 1$ **to** $T$ **do**
8     **for** $j \leftarrow 1$ **to** $N \times \Psi$ **do**
9         Sample with replacement $\theta_j^*$ from the weighted set $\{\theta_i^{(t-1)}, w_i^{(t-1)}\}_{i=1,\ldots,N}$;
10         Perturb $\theta_j^*$: $\theta_j^{**} \sim \mathcal{K}_{t-1}(\cdot \mid \theta^*)$;
11         if $\pi(\theta_j^{**}) = 0$ go to line 10, otherwise continue;
12         Draw synthetic sample $\mathbf{x}_j^{**}$ from the model $f(\cdot \mid \theta_j^{**})$;
13         Compute the summary statistics $\eta(\mathbf{x}_j^{**})$;
14         Compute the distance $\Delta_j = d(\eta(\mathbf{x}_j^{**}), \eta(\mathbf{x}))$;
15     **end**
16     Let $\Delta_{(1)} \leq \Delta_{(2)} \leq \cdots \leq \Delta_{(N \times \Psi)}$ be the sorted distances;
17     Set the new population $\{\theta_i^{(t)}\}_{i=1}^N$ to be the particles $\theta_j^{**}$ for which $\Delta_j \leq \Delta_{(N)}$;
18     Set $\tilde{w}_i^{(t)} = \pi(\theta_i^{(t)}) / \sum_{j=1}^N w_j^{(t-1)} \mathcal{K}_{t-1}(\theta_i^{(t)} \mid \theta_j^{(t-1)})$;
19     Normalize the weights $w_i^{(t)} := \tilde{w}_i^{(t)} / \sum_{j=1}^N \tilde{w}_j^{(t)}$
20 **end**

## E.2 ABC-CNN

The approach of Akesson et al. [2022] is simply a small modification of the standard ABC, where CNNs are used to estimate the posterior mean of a parameter, which is then used as a summary statistic. The line 8 of the standard ABC algorithm (Algorithm 1 in the article) to compute summary statistics, is replaced by the prediction of a CNN trained on the reference table, see lines 11, 14 and 15 of Algorithm 4. The output of the algorithm is similar to the one of the standard ABC: for a given data sample $\mathbf{x}$ and a tolerance threshold, we hope that the parameters $\theta$ accepted during the process are approximately distributed from the posterior distribution $\pi(\theta \mid \mathbf{x})$. Hence, to have asymptotic consistency and valid frequentist coverages, the same kind of conditions are needed than for standard ABC (Algorithm 1 in the article). But it's harder to check these conditions for ABC-CNN than for standard ABC, because the summary statistic used in the ABC-CNN is the output of a CNN which is a "black-box". It is therefore practically impossible to verify anything about this summary statistic.

Although, there are some important differences that are worth highlighting between the two methods. The main and most important one is that it is not necessary to give summary statistics, thanks to the use of a CNN which automatically generates a relevant one. However, a distance $d$ for comparing two data samples and a tolerance threshold $\alpha$ are still required. In practice, this distance is often simpler to determine in the case of ABC-CNN than in the case of standard ABC, because comparing two parameters is often simpler than comparing two sets of summary statistics, which can be numerous and very diverse. Theoretically, as explained by Frazier et al. [2018], for a fixed choice of summaries, the two-stage procedure advocated by Fearnhead and Prangle [2012] and used by Akesson et al. [2022], will not reduce the asymptotic variance over a point estimate of a parameter produced via staandard ABC (Algorithm 1 in the article). However, this two-stage procedure does reduce the Monte Carlo error inherent in estimating the approximate posterior distribution $\pi_{\alpha,d}(. \mid \eta(\mathbf{x}))$ by reducing the dimension of the statistics on which the matching in approximate Bayesian computation is based. Using this approach, we can hope to have a smaller global error of approximation of the true posterior, compared to standard ABC.

Moreover, a CNN can deal with different types of data, in particular high dimensional and complex data, like temporal or spatial data. Parameters can be easily estimated from these complex data by a CNN, while it can be difficult to find appropriate summary statistics to be used in a classical ABC.

Another difference with the standard ABC is that a validation dataset should be generated, to be used to choose the network architecture, as usual for neural networks. This dataset is generally much smaller in size than the reference table used as a training set. Therefore, it generally adds little computation time compared with the effort required to generate the reference table. This lost computation time can be recovered, and time can even be saved, in the classical ABC step. Indeed, in this step, the sample of interest $\mathbf{x}$ for which we want to approximate the posterior $\pi(\theta \mid \mathbf{x})$ should be compared with all the samples in the reference table (lines 9 of Algorithm 1 and 15 of Algorithm 4). This comparison can be time-consuming, depending on the distance used between summary statistics and the

**Algorithm 4:** Pseudocode for CNN as Summary Statistics for ABC [Akesson et al., 2022]

**Input :** A Bayesian parametric model $\{f(\cdot \mid \theta), \pi\}$, a data sample $\mathbf{x}$, integers $N_{\text{train}}$ and $N_{val}$ representing sizes of training, and validation sets, a (pseudo)-distance $d$ between two samples, and a tolerance threshold $\alpha \in (0, 1]$.

**Output:** A set $\theta_1, \ldots, \theta_{[N_{\text{train}}\alpha]}$ whose distribution is approximately $\pi(\theta \mid \mathbf{x})$

1 **Generation of a reference table (training dataset)** :
2 **for** $j \leftarrow 1$ **to** $N_{\text{train}}$ **do**
3 | Draw $\theta_j \sim \pi$;
4 | Draw synthetic sample $\mathbf{x}_j = (x_{1,j}, \ldots, x_{d,j})^\top$ from the model $f(\cdot \mid \theta_j)$;
5 **end**
6 **Generation of a validation dataset** :
7 **for** $j \leftarrow 1$ **to** $N_{val}$ **do**
8 | Draw $\theta_j \sim \pi$;
9 | Draw synthetic sample $\mathbf{x}_j = (x_{1,j}, \ldots, x_{d,j})^\top$ from the model $f(\cdot \mid \theta_j)$;
10 **end**
11 **Train a CNN on the reference table**: the training pairs are the $\{(\mathbf{x}_j, \theta_j), j = 1, \ldots, N_{\text{train}}\}$. The validation set is used to choose the best architecture of the CNN.
12 **Summary statistics and distances** :
13 **for** $j \leftarrow 1$ **to** $N_{\text{train}}$ **do**
14 | $\mathbf{x}_j$ is given as input to the trained CNN to obtain $\eta_{\text{CNN}}(\mathbf{x}_j)$ ;
15 | Compute the distance $d_j = d(\eta_{\text{CNN}}(\mathbf{x}), \eta_{\text{CNN}}(\mathbf{x}_j))$ ;
16 **end**
17 Order these distances, i.e., $d_{(1)} < d_{(2)} < \cdots < d_{(N_{\text{train}})}$;
18 Keep the $\theta_j$ corresponding to the $\lfloor N_{\text{train}}\alpha \rfloor$ smallest distances.

size of the reference table. If there are many summary statistics to be used by the standard ABC, the ABC-CNN which uses only one summary statistic, will be much faster. The larger the reference table, the greater the difference. This is why, in many cases, ABC-CNN will often be faster than standard ABC, despite the generation of a validation set. However, we can not generalize, and it clearly depends on the problem at hand.

## E.3 ABC-RF

Raynal et al. [2018] propose to conduct likelihood-free Bayesian inferences about parameters with no prior selection of the relevant components of summary statistics and bypassing the derivation of the associated tolerance threshold. Their approach relies on the random forest (RF) methodology of Breiman [2001] applied in a regression setting.

---

**Algorithm 5:** Pseudocode for an Approximate Bayesian Computation - Random Forest sampler

---

**Input :** A Bayesian parametric model $\{f(\cdot \mid \theta), \pi\}$, a data sample $\mathbf{x}$, $N_{\text{train}}$ the size of the training set, summary statistics $\eta : \mathbb{R}^d \to \mathbb{R}^l$.

**Output:** For $T(\theta)$ a scalar transform of $\theta$, approximations of the posterior expected value $\psi_\eta(\mathbf{x}) = \mathbb{E}_\pi[T(\theta) \mid \eta(\mathbf{x})]$, the posterior variance $\mathbb{V}_\pi[T(\theta) \mid \eta(\mathbf{x})]$, and posterior quantiles.

1 **Generation of a reference table** :
2 **for** $j \leftarrow 1$ **to** $N_{\text{train}}$ **do**
3 | Draw $\theta_j \sim \pi$;
4 | Draw synthetic sample $\mathbf{x}_j = (x_{1,j}, \ldots, x_{d,j})^\top$ from the model $f(\cdot \mid \theta_j)$
5 **end**
6 **Summary statistics** :
7 **for** $j \leftarrow 1$ **to** $N_{\text{train}}$ **do**
8 | Compute the vector of summary statistics $\eta(\mathbf{x}_j)$
9 **end**
10 **Train a random forest**: the training pairs are the $\{(\mathbf{x}_j, \theta_j), j = 1, \ldots, N_{\text{train}}\}$.
11 **Approximate** $\psi_\eta(\mathbf{x})$: give $\eta(x)$ as input of the trained random forest: the output of the RF allows to obtain easily approximations of $\psi_\eta(\mathbf{x}) = \mathbb{E}_\pi[T(\theta) \mid \eta(\mathbf{x})]$, $\mathbb{V}_\pi[T(\theta) \mid \eta(\mathbf{x})]$ and of posterior quantiles (hence credible intervals).

---

This algorithm automatizes the inclusion of summary statistics and do not need the definition of a distance or the choice of a tolerance threshold. It appears mostly insensitive to the presence of non-relevant summary statistics. It can then deal with a large number of summary statistics, by-passing any form of pre-selection. However, it is mandatory that relevant ones are present in the initial pool of summary statistics to be considered. To obtain posterior quantiles and then credible intervals, this ABC-RF method uses quantile regression forests to approximate the posterior cumulative distribution function of $T(\theta)$ given $\eta(\mathbf{x})$, $F(T(\theta) \mid \eta(\mathbf{x}))$. The asymptotic consistency of this approach has been established by Meinshausen [2006], under some conditions. In particular, $F(T(\theta) \mid \eta(\mathbf{x}))$ should be Lipschitz continuous and strictly monotonously increasing in $T(\theta)$. Such conditions are practically unverifiable in an ABC framework where the likelihood is unknown or intractable.

In practice, Raynal et al. [2018] showed on several examples that the approximations of posterior expectations obtained by ABC-RF were quite accurate, while posterior variances were slightly overestimated, and confidence sets slightly conservative (larger than the exact ones).

# F Detailed comparison of the five ABC methods used in the examples

Below the five methods used (standard ABC, ABC-SMC, ABC-CNN, ANC-RF and ABCD-Conformal) are compared in details on different points, with a practical point of view. Table 6 in the article summarizes these reflections.

**Output of the algorithm**
First, one big difference between these methods is what is obtained from the algorithm: for standard ABC, ABC-SMC and ABC-CNN an approximation of the whole posterior distribution of the parameter of interest is obtained. The ABC-RF and ABCD-Conformal methods focus on the approximation of transforms of interest of the posterior, like posterior mean, posterior variance or posterior quantiles for instance. It is rather common that only transforms of the posterior are of interest for practitioners, and in this case all four methods can be used. But if the whole posterior would be studied, ABC-RF and ABCD-Conformal will not be adapted. Note that to obtain a good approximation of the whole posterior, in general more simulations are required than for simply estimating transforms of the posterior.

**Need of relevant summary statistics**
Standard ABC and ABC-SMC need summary statistics to be able to compare simulations and observed data. Sometimes obvious relevant statistics exist (like in the MA2 example for instance), and the best is to have exhaustive statistics. But most of the time, it is not easy to find such relevant statistics, and it is a serious bottleneck when performing inference on complex and high-dimensional problems. As explained in section 2.1 of the article, it is preferable to have a small number of summary statistics to avoid the burden of multidimensionality, and these statistics should be sufficiently informative [Fearnhead and Prangle, 2012]. ABC-RF enables to automatize the inclusion of summary statistics in an ABC algorithm. But it is necessary to have a set of statistics from which to select relevant ones: relevant statistics should be in the set, and this is a limitation. ABC-CNN and ABCD-Conformal do not need any summary statistic, which is an advantage, as the choice of these summaries have been proven to be crucial to obtaining good results.

**Need for a distance**
A distance to compare simulations and observed data is needed for standard ABC and ABC-SMC. This distance depends on the summary statistics used and on the problem at hand. Specific distances can be used for genetics problem, for instance. In general, a good knowledge of the problem in question and discussions with experts are necessary to define a relevant distance. In case of low-dimensional statistics, relevant distances are generally faster to compute. But in case of high-dimensional statistics, they can be quite long to compute. This is the case of the Lotka-Volterra example: we did not really use summary statistics, as datasets and simulations were directly compared using an Euclidian distance. But as the dataset was of dimension forty (two series of twenty times), it was computationally demanding: the standard ABC was the method which took more computing time. Regarding the ABC-CNN algorithm, a summary statistic is obtained through the network, and the distance to be used is in general quite simple and fast to be computed. Of course computing time increases with the dimension of the parameters of interest. ABC-RF and ABCD-Conformal do not need any distance, which can avoid a tricky choice and can save computational time.

**Need for a tolerance threshold**

For standard ABC and ABC-CNN, a tolerance threshold is needed, and its choice affects the degree of approximation obtained for the whole distribution. Usually, for practical reasons, quantile-based acceptance thresholds are used (quantiles of the distances between the observed data set and simulations in the training set). ABC-SMC requires a decreasing sequence of tolerance levels, or at least a first threshold and a way to decrease it between each population (usually using percentiles). Like for standard ABC and ABC-CNN, the choice of this sequence can affect a lot the degree of approximation obtained. On the opposite, ABC-RF and ABCD-Conformal do not need a tolerance threshold, as they do not use a distance.

**Dealing with multidimensional parameters**

Concerning the parameter of interest, the ABC-RF method can estimate only uni-dimensional transforms of interest, usually a projection on a given coordinate of the parameter. In the discussion of their article, Raynal et al. [2018] explained that their attempts to deal with multidimensional parameters were so far unfruitful. The standard ABC and the ABC-SMC can deal with multidimensional parameter, but it suffers from the burden of multidimensionality, as summary statistics and distances are more difficult to define, and take more time to be computed in high dimension. ABC-CNN, despite having a standard ABC step in the algorithm, suffers less from this burden, as the summary statistic obtained through the CNN is of the same dimension as the parameter of interest. For ABCD-Conformal, multidimensionality is not a problem, as the computing time for the CNN with Dropout and the conformal procedure does not really increase with the dimension (at least on the examples studied).

**Choose network or random forest architecture**

ABC-CNN and ABCD-Conformal are based on neural networks. Hence, the architecture of the networks to be used should be chosen: mainly the number of layers, the type of layer, the number of neurons in a layer, the size of the kernel for convolutional layers, and the activation functions. For the Dropout rate, we do not need to choose it when using the concrete Dropout approach of Gal et al. [2017]. In practice, for the examples studied, we did not spend too much time to choose the architecture, as it is quite fast to train different networks and to compare them on the validation set. The most influential parameters for us in our examples were the type of layers (convolutional or dense), and the activation functions. Similarly, to use ABC-RF some parameters should be chosen for the random forest, mainly the number of trees, the minimum node size and the proportion of summary statistics sampled at each split. For this method also, we did not encounter difficulties to set these parameters, standard values making the job on our example.

**Needed datasets**
Standard ABC and ABC-RF need a training set of simulations, to be compared with the observed data. ABC-CNN also needs a validation set, which is used to choose the network architecture. The ABCD-Conformal also needs a calibration set for the conformal procedure. This calibration set is usually of size 500 or 1000 depending on the accuracy we want for the marginal coverage [Angelopoulos and Bates, 2023]. Therefore, more simulations are needed for ABC-CNN than for standard ABC and ABC-RF, and even more for ABCD-Conformal. However, the sizes of the validation and of the calibration sets are generally quite smaller than the size of the training set, and the increase in simulation time is small compared to what is required for the training set. This time can often be recovered in other steps of the algorithm. Concerning the ABC-SMC, it cannot rely on datasets generated in advance. Simulations must be performed sequentially, and are specific to each data sample of interest. In practice, ABC-SMC therefore requires significantly more time for calibration and testing compared to the other methods.

**Theoretical justification of the methods**
In this article, we are particularly interested in the estimation of parameters of interest, with associated confidence sets. The question of the frequentist coverage of the obtained intervals matters, as the intervals represent our uncertainty of our estimation. We then compared the frequentist coverage of the obtained intervals for the different methods, on four examples, and we studied the bibliography for the theoretical justification of all these methods.

As detailed in section 2.1 of the article, Frazier et al. [2018] studied asymptotic properties of standard ABC and gave conditions under which Bayesian confidence sets have valid frequentist coverage levels. Hence, in theory and asymptotically, under some regularity conditions, standard ABC and ABC-CNN methods enable posterior concentration and the *probability matching criterion*. But these conditions can not be verified in general, and this restricts the use and validity of these methods.

Concerning ABC-SMC, the convergence of importance sampling-based algorithms has been extensively studied, as many different algorithms have been proposed, see for instance Beaumont et al. [2009]; Del Moral et al. [2012]. Convergence requires specific conditions, in particular regarding the sequence of tolerance thresholds, and the choice of perturbation samplers (transition kernels).

As said in section 2.3.2 in the article, ABC-RF is based on quantile regression forests whose asymptotic consistency has been proved by Meinshausen [2006] under some conditions. But these conditions are practically unverifiable in an ABC framework.

On the opposite, the ABCD-Conformal method guarantees a non-asymptotic marginal frequentist coverage, without any distributional or model assumption. The test samples should solely come from the same distribution as the calibration samples. Only the efficiency of the conformal procedure is in question, as the lengths of the confidence sets will depend on the quality of the uncertainty measure used in the procedure. This guaranteed coverage is a great advantage compared to the other methods. However, this coverage is guaranteed marginally, and not conditionally. Nevertheless, some metrics can be computed to check how close we are to conditional marginality, see Angelopoulos and Bates [2023].

**Benefits of using concrete Dropout**

Apart from the fact that epistemic and aleatoric uncertainties are used to obtain confidence intervals during the conformal procedure, their values themselves and the distinction between these uncertainties are interesting for effective exploration of the uncertainty: which part can be attributed to the model or to the noise in the data, in which circumstances?